\documentstyle[11pt]{article}
\def\setfonts{%
\font\openbig=msbm10 scaled\magstephalf
\font\openscr=msbm8 
\font\openscrscr=msbm8
\newfam\openfam
\textfont\openfam=\openbig
\scriptfont\openfam=\openscr
\scriptscriptfont\openfam=\openscrscr
\def\open{\fam\openfam}

\font\ssfbig=cmss10 scaled\magstephalf
\font\ssfscr=cmss8 
\font\ssfscrscr=cmss8
\newfam\ssffam
\textfont\ssffam=\ssfbig
\scriptfont\ssffam=\ssfscr
\scriptscriptfont\ssffam=\ssfscrscr
\def\ssf{\fam\ssffam}
}

\makeatletter
\@addtoreset{equation}{section}
\newdimen\normalarrayskip
\newdimen\minarrayskip
\normalarrayskip\baselineskip
\minarrayskip\jot
\newif\ifold \oldtrue \def\new{\oldfalse}
\def\arraymode{\ifold\relax\else\displaystyle\fi}

\def\@arrayskip{\ifold\baselineskip\z@\lineskip\z@
  \else
  \baselineskip\minarrayskip\lineskip2\minarrayskip\fi}
\def\@arrayclassz{\ifcase \@lastchclass \@acolampacol \or
\@ampacol \or \or \or \@addamp \or
 \@acolampacol \or \@firstampfalse \@acol \fi
\edef\@preamble{\@preamble
 \ifcase \@chnum
  \hfil$\relax\arraymode\@sharp$\hfil
  \or $\relax\arraymode\@sharp$\hfil
  \or \hfil$\relax\arraymode\@sharp$\fi}}
\def\@array[#1]#2{\setbox\@arstrutbox=\hbox{\vrule
  height\arraystretch \ht\strutbox
  depth\arraystretch \dp\strutbox
  width\z@}\@mkpream{#2}\edef\@preamble{\halign \noexpand\@halignto
\bgroup \tabskip\z@ \@arstrut \@preamble \tabskip\z@ \cr}%
\let\@startpbox\@@startpbox \let\@endpbox\@@endpbox
 \if #1t\vtop \else \if#1b\vbox \else \vcenter \fi\fi
 \bgroup \let\par\relax
 \let\@sharp##\let\protect\relax
 \@arrayskip\@preamble}
\@addtoreset{equation}{section}
\newcounter{@sc}
\newcounter{@scp}
\newcounter{@t}
\newlength{\@x}
\newlength{\@xa}
\newlength{\@xb}
\newlength{\@y}
\newlength{\@ya}
\newlength{\@yb}
\newsavebox{\@pt}

\def\bezier#1(#2,#3)(#4,#5)(#6,#7){\c@@sc#1\relax
  \c@@scp\c@@sc \advance\c@@scp\@ne
  \@xb #4\unitlength \advance\@xb -#2\unitlength \multiply\@xb \tw@
  \@xa #6\unitlength \advance\@xa -#2\unitlength
      \advance\@xa -\@xb \divide\@xa\c@@sc
  \@yb #5\unitlength \advance\@yb -#3\unitlength \multiply\@yb \tw@
  \@ya #7\unitlength \advance\@ya -#3\unitlength
      \advance\@ya -\@yb \divide\@ya\c@@sc
  \setbox\@pt\hbox{\vrule height\@halfwidth  depth\@halfwidth
   width\@wholewidth}\c@@t\z@
   \put(#2,#3){\@whilenum{\c@@t<\c@@scp}\do
      {\@x\c@@t\@xa \advance\@x\@xb \divide\@x\c@@sc \multiply\@x\c@@t
       \@y\c@@t\@ya \advance\@y\@yb \divide\@y\c@@sc \multiply\@y\c@@t
       \raise \@y \hbox to \z@{\hskip \@x\unhcopy\@pt\hss}%
       \advance\c@@t\@ne}}}
\makeatother

\def\theequation{\thesection.\arabic{equation}}
\setfonts

\def\lvm{\leavevmode\hbox to\parindent{\hfill}}
\def\req#1{(\ref{#1})}

\def\BE{\begin{equation}}
\def\EE{\end{equation} }
\def\BA{\begin{array}} 
\def\EA{\end{array}}
\def\L{\left}
\def\R{\right}

\def\frac#1#2{{\textstyle{{#1}\over{#2}}}}
\def\ket#1{\bigl|{#1}\bigr\rangle}
\def\nket#1{|{#1}\rangle}
\def\d{\partial}

\def\N#1{N\!=\!#1}
\def\SL#1{s\ell(#1)}

\def\half{{\textstyle{1\over2}}}

\def\cA{{\cal A}}

\def\cG{{\cal G}}
\def\cH{{\cal H}}

\def\cL{{\cal L}}

\def\cP{{\cal P}}
\def\cQ{{\cal Q}}

\def\cT{{\cal T}}

\def\cV{{\cal V}}

\def\oC{{\open C}}
\def\oN{{\open N}}
\def\oQ{{\open Q}}
\def\oZ{{\open Z}}

\def\ctop{{\ssf c}}
\def\htop{{\ssf h}}

\def\tensor{\otimes}
\def\tilde{\widetilde}

\def\NPB{Nucl.\ Phys.\ B}
\def\PLB{Phys.\ Lett.\ B}
\def\MPLA{Mod.\ Phys.\ Lett.\ A}

\def\IJMPA{Int.\ J.\ Mod.\ Phys.\ A}

\begin{document}
\hfuzz=1.8pt
\begin{flushright}
{\tt hep-th@xxx/9512079}
\end{flushright}
\thispagestyle{empty}

\begin{center}
{\Large{\sc Singular Vectors of the Topological Conformal Algebra}}\\[6pt]
{\large A.~M.~Semikhatov and I.~Yu.~Tipunin}\\[6pt]
{\small\sl I.~E.~Tamm Theory Division,
P.~N.~Lebedev Physics Institute, Russian Academy of Sciences
}
\end{center}
\vskip-4pt
\addtolength{\baselineskip}{-2pt}
{\footnotesize
A general construction is found for `topological' singular vectors of the
twisted $\N2$ superconformal algebra.  It demonstrates many parallels with
the known construction for affine $\SL2$ singular vectors due to
Malikov--Feigin--Fuchs, but is formulated independently of the latter. The
two constructions taken together provide an isomorphism between the
topological and affine $\SL2$ singular vectors. The general formula for
topological singular vectors can be reformulated as a chain of direct
recursion relations that allow one to derive a given singular vector
$\nket{S(r,s)}$ from the lower ones $\nket{S(r,s'<s)}$.  We also introduce
generalized Verma modules over the twisted $\N2$ algebra and show that they
provide a natural setup for the new construction for topological singular
vectors.}

\addtolength{\baselineskip}{2pt}

\medskip

\section{Introduction}\lvm
Singular vectors in Verma modules of infinite-dimensional algebras
represented in conformal models play an important role both in conformal
field theory and in the representation theory.  Their significance in
conformal models was appreciated already in~\cite{[BPZ]}, where decoupling
conditions for singular vectors were shown to lead to differential equations
on correlators.  Mathematically, this is a manifestation of the fact that
irreducible representations are obtained by factoring the Verma modules over
submodules built on singular vectors~\cite{[FF]}.  `Positions' of singular
vectors determine embeddings of Verma modules. A number of papers have been
devoted to finding positions as well as explicit representations, and various
ways to derive, singular vectors in different models
~\cite{[KK]}--\cite{[Doerr]}.

A closed formula for all singular vectors in Verma modules over $\SL{n}$ (in
particular, $\SL2$) Ka\v c--Moody algebras was found in~\cite{[MFF]}, and it
has stimulated efforts aimed at deriving singular vectors for other popular
algebras (Virasoro, $\N2$, $W_3$, $\ldots$) related to affine Lie algebras.
Morphisms between [conformal field theories with] different algebras  have
been used to find mappings between the corresponding singular states.  For
instance, hamiltonian reduction of $\SL2$ and $\SL3$ affine algebras allows
one to reduce the corresponding Ka\v c--Moody singular vectors to those in
the appropriate `matter' theories.  These reductions, however, do not provide
{\it intrinsic\/} constructions for singular vectors in algebras other than
the affine Lie algebras.  A challenging problem is to find generating
constructions for singular vectors using the idea of `analytic' (in fact,
{\it algebraic\/}) continuation developed in terms of a given algebra only.

The aim of the present paper is to give a complete realization of such a
program for the so called {\sl topological\/} singular vectors, i.e.\ BRST-
($\cQ_0$-) or $\cG_0$- invariant singular vectors of the twisted $\N2$
superconformal algebra, built upon BRST-invariant {\it chiral\/} primary
states.  We will present a closed algebraic construction for all topological
singular vectors in Verma modules.  This construction is in fact a particular
case of a more general one that produces singular vectors in {\it
generalized\/} Verma modules.

It will be seen that the proposed construction has many similarities with the
MFF one, although it relies on more involved algebraic properties.  It is
indeed motivated by the results of~\cite{[S-sing]}, which strongly suggest an
isomorphism between $\SL2$- and topological singular vectors.  One should
recall the apparent simplicity of the MFF formula, in which an involved
algebraic expression is `packed' into a single {\it monomial\/}; that
monomial, however, does not look like an element of a Verma module, and a set
of algebraic transformations is required in order to bring it to the `Verma'
form.  It is in the course of this rearrangement that complexity of the
expression grows rapidly.

Just like the MFF formula, the general formula for topological singular
vectors (eqs.~\req{N2sing} and \req{Tminus}) looks deceptively simple, but it
also does require certain algebraic manipulations in order to be transformed
into a `Verma' form.  It involves objects that look like a continuation of
products of modes of the fermionic operators $\cG$ and $\cQ$ ($G$ and
$\overline G$ or $G^\pm$ in a different nomenclature) to a non-integral
number of factors.  These objects will be called
intertwining\,\footnote{Strictly speaking, these operators do not commute
with the action of the algebra and thus are not intertwiners, but they can be
realized as such by integral operators on the appropriate supermanifold in
the spirit of \cite{[FM]}.  We will therefore use the convenient name
`intertwiners' for these operators in the future.\label{foot:intertw}}
operators; they map between modules related by the spectral flow
transform~\cite{[SS],[LVW]}.  The appearance of the intertwining operators is
very natural since any singular vector in a given module can be viewed as an
intertwiner of that module with itself. We arrive at the general construction
for topological singular vectors by simply looking for the appropriate
intertwiners\,\footnote{In the MFF case~\cite{[MFF]}, the `intertwiners' can
be written as $(J^+_0)^p$ and $(J^-_{-1})^p$, $p\in\oC$, which makes some of
their algebraic properties self-evident. In the $\N2$ case, more relations
will have to be formulated explicitly, which is one of the reasons for the
construction to appear more involved.}. Then, to compute a singular vector in
the Verma form, we will need a set of commutation relations between the
intertwiners and the generators of the topological conformal algebra.

A very useful reformulation of the construction for topological singular
vectors can be given in the form of a sequence of recursion relations
between topological singular vectors of different levels.  That is, given a
singular vector $\nket{S(r,s)}^\pm$, there is a unique procedure of building
the singular vector $\nket{S(r,s+1)}^\mp$.  Our construction can thus be
reformulated as a method to derive a given topological singular vector from a
`lower' one.  Such a recursive reformulation is very convenient for
computational purposes, but it can also be used to prove by induction
that the construction does indeed give singular vectors.  In whatever
reformulation, the proposed construction is {\sl direct\/} in that it does
not require solving `inverse' problems such as e.g.\ finding a solution of
(however simple) system of equations.

\medskip

We begin in the next section with setting up the problem and then recall in
section \ref{sec:MFF} the MFF formula for the affine $\SL2$ singular vectors
and the Kazama--Suzuki mapping. Analogies with the $\SL2$ case  are quite
useful in motivating the subsequent construction, but the properties of the
intertwiners necessary in the topological case have to be formulated
independently, which we do in section~\ref{sec:Inter}.  Then, in section
\ref{sec:Gen}, we give the general formula for topological singular vectors.
The algebraic scheme that allows one to rewrite these as elements of the
usual Verma modules is described in section \ref{sec:Back}.  The recursive
reformulation is given in section~\ref{sec:Rec}, where we explain how
topological singular vectors on different levels are related by purely
algebraic manipulations.  Section \ref{sec:Concl} contains several
concluding~remarks.

The main results of this paper have been announced in~\cite{[ST2r]}.

\section{Topological conformal algebra and singular vectors}\lvm
By the topological [conformal] algebra (see~\cite{[EY],[W-top]}) we will mean
the twisted $\N2$ superconformal algebra spanned by the integer-moded
generators $\cL_m$ (Virasoro generators), $\cQ_m$ (BRST current), $\cH_m$
($U(1)$ current), and $\cG_m$ (fermionic spin-2 current):
\BE\new\BA{lclclcl}
\L[\cL_m,\cL_n\R]&=&(m-n)\cL_{m+n}\,,&\qquad&[\cH_m,\cH_n]&=
&{\ctop\over3}m\delta_{m+n,0}\,,\\
\L[\cL_m,\cG_n\R]&=&(m-n)\cG_{m+n}\,,&\qquad&[\cH_m,\cG_n]&=&\cG_{m+n}\,,
\\
\L[\cL_m,\cQ_n\R]&=&-n\cQ_{m+n}\,,&\qquad&[\cH_m,\cQ_n]&=&-\cQ_{m+n}\,,\\
\L[\cL_m,\cH_n\R]&=&\multicolumn{5}{l}{-n\cH_{m+n}+{\ctop\over6}(m^2+m)
\delta_{m+n,0}\,,}\\
\L\{\cG_m,\cQ_n\R\}&=&\multicolumn{5}{l}{2\cL_{m+n}-2n\cH_{m+n}+
{\ctop\over3}(m^2+m)\delta_{m+n,0}\,,}\EA\qquad m,~n\in\oZ\,.
\label{topalgebra}
\EE

The singular vectors we are interested in are built on `topological primary
states'$\!$, which are defined by a set of annihilation conditions (to be
called the `highest-weight' conditions for brevity) that can be described as
follows.  First, one imposes the conditions
\BE
\cQ_{\geq0}\ket\htop_{\rm top}=\cG_{\geq1}\ket\htop_{\rm top}=
\cL_{\geq1}\ket\htop_{\rm top}=
\cH_{\geq1}\ket\htop_{\rm top}=0
\label{upper}\EE
(the $\cQ_0$-condition is sometimes referred to as BRST-invariance). In
addition, one imposes chirality~\cite{[LVW]}, which in the present, twisted,
version amounts to
\BE
\cG_{0}\ket\htop_{\rm top}=0\,.
\label{G0}\EE
This implies then that the eigenvalue of $\cL_0$ vanishes, hence the `Cartan'
part of the highest-weight conditions reads
\BE
\cL_{0}\ket\htop_{\rm top}=0\,,\qquad
\cH_0\ket\htop_{\rm top}=\htop\ket\htop_{\rm top}
\label{Cartan}\EE
Equations \req{upper}--\req{Cartan} determine what we will call the topological
primary states; $\htop$ will be called the topological $U(1)$ charge.

A BRST-invariant topological singular vector built on a state $\ket\htop_{\rm
top}$ is required to satisfy the same annihilation conditions as
in~\req{upper},
\BE
\cQ_{\geq0}\ket{S}^+=\cG_{\geq1}\ket{S}^+=
\cL_{\geq1}\ket{S}^+=
\cH_{\geq1}\ket{S}^+=0\,.
\EE
This determines a half of singular vectors that can be built on the
topological primary states, the other half being determined by
\BE
\cQ_{\geq1}\ket{S}^-=\cG_{\geq0}\ket{S}^-=
\cL_{\geq1}\ket{S}^-=
\cH_{\geq1}\ket{S}^-=0\,.
\EE

These singular vectors exist in a Verma module $V_\htop$ over
$\nket\htop_{\rm top}$ if
$\htop=\htop^+(r,s)$ or $\htop=\htop^-(r,s)$ respectively
\cite{[BFK],[S-sing]}, where
\BE\new\BA{rcl}
\htop^+(r,s)&=&-\frac{r-1}{t}+s-1\,,\\
\htop^-(r,s)&=&\frac{r+1}{t}-s\,,\EA\qquad r,s\in\oN
\label{htoprs}\EE
and $t\equiv k+2$ is related to the topological central charge $\ctop$ by
$\ctop=3(t-2)/t$.  Our aim is to find an explicit formula for these singular
vectors for arbitrary $r$ and $s\in\oN$.

Denote by $V^+_{r,s}$ and $V^-_{r,s}$ respectively the Verma modules built
on the topological primary states with topological $U(1)$
charges~\req{htoprs}.  We will mainly concentrate for definiteness on the
singular vectors, denoted as $\nket{S(r,s)}^+$, in the `$+$'-modules
$V^+_{r,s}$. Let us note right now that all such vectors have the structure
\BE\nket{S(r,s)}^+ =
\cQ_0\ldots\cQ_{r-1}\cdot\nket{T(r,s)}^+\,,
\label{ST}
\EE
where $\nket{T(r,s)}^+$ is a Verma module element that, too, satisfies
certain `highest-weight' conditions
\BE\new\BA{l}
\cQ_{\geq r}\ket{T(r,s)}^+=\cG_{\geq -r}\ket{T(r,s)}^+=
\cL_{\geq1}\ket{T(r,s)}^+=
\cH_{\geq1}\ket{T(r,s)}^+=0\,,\\
\cH_{0}\ket{T(r,s)}^+=(\htop^+(r,s)+r)\ket{T(r,s)}^+\,.
\EA\EE
In the `${}-{}$'-modules, similarly,
\BE\nket{S(r,s)}^-=
\cG_0\ldots\cG_{r-1}\cdot\nket{T(r,s)}^-
\EE
The vectors \ $\nket{S(r,s)}^\pm$ are on level $rs$ over
$\nket{\htop^\pm(r,s)}_{\rm top}$.  The non-trivial part of the construction
is finding these $\nket{T(r,s)}^\pm$. The general formulae will be given
in~\req{N2sing} and~\req{Tminus}, and a large part of
sections~\ref{sec:Inter}--\ref{sec:Rec} will be devoted to explaining their
meaning.

\section{Reminder on the MFF construction and related
formulae\label{sec:MFF}}\lvm
A motivation for our construction can be taken partly from the Kazama--Suzuki
mapping~\cite{[KS]}, which allows one to map the topological conformal
algebra to the universal enveloping of $\SL2_k\oplus[BC]$, where $[BC]$ is an
auxiliary fermionic $BC$-system (spin-1 `ghosts'), and $\SL2_k$ is a
level-$k$ Ka\v c--Moody algebra:
\begin{eqnarray} &&\cQ=CJ^+\,,\quad
\cG=\frac{1}{k+2}BJ^-\,,\quad
\cH=\frac{k}{k+2}BC-\frac{2}{k+2}J^0\,,\label{GQHks}\\
&&\cT=\frac{1}{k+2}(J^+J^-)-\frac{k}{k+2}B\d
C-\frac{2}{k+2}BCJ^0\label{Tks}
\end{eqnarray}
Topological central charge $\ctop$ of the topological generators thus
constructed is given by $\ctop={3k\over k+2}$.

As ref.~\cite{[S-sing]} indicates, the Kazama--Suzuki mapping induces an
isomorphism between the topological and affine $\SL2$ singular
vectors\footnote{Considered in~\cite{[S-sing]} were only the $\SL2$-singular
vectors over the highest-weight state $\nket{j_+(r,s)}$ (see eq.~\req{jplus}
below).  A similar statement holds for the `other half' of singular vectors
$\nket{{\rm MFF\/}\{r,s\}}^-$, namely those built upon
$\nket{j_-(r,s)}$ where $j_-(r,s)=-\half(r+1)+\half(k+2)s$ and $r,s\in\oN$.}:
evaluating the image of a topological singular vector $\nket{T(r,s)}^+$ under
the Kazama--Suzuki mapping \req{GQHks}, \req{Tks}, we find that it literally
coincides with an $\SL2$ singular vector:
\BE
\ket{T(r,s)}^+\mapsto
\ket{{\rm MFF}\{r,s\}}^+\tensor\ket0_{BC}
\label{isomorphism}
\EE
where $\nket0_{BC}$ is the $BC$ vacuum, the vacuum annihilation conditions
being \ $B_{\geq0}\nket0=C_{\geq1}\nket0=0$, and $\nket{{\rm MFF}\{r,s\}}^+$
is a singular vector in the affine $\SL2_k$ Verma module with the
highest-weight state $\ket{j_+(r,s)}_{\SL2}$, where
\BE
j_+(r,s)=\frac{r-1}{2} - t\frac{s-1}{2}\,,\qquad r,s\in\oN\,,\qquad
t\equiv k+2\,.
\label{jplus}
\EE

The formula~\req{isomorphism}, although it has been thoroughly established
only for level $\leq4$, does nevertheless strongly suggest a 1:1 isomorphism
between topological and $\SL2$ singular vectors. A general formula is known
for the latter~\cite{[MFF]}~\footnote{Again, we will consider for brevity
only a `half' of the MFF singular vectors, the $\nket~^+$-ones.}:
\BE\ket{{\rm MFF}\{r,s\}}=
(J^-_0)^{r+(s-1)t}(J^+_{-1})^{r+(s-2)t}(J^-_0)^{r+(s-3)t}\ldots
(J^+_{-1})^{r-(s-2)t}
(J^-_0)^{r-(s-1)t}\ket{j_+(r,s)}_{\SL2}\,.\label{mff}\EE
These will thus be called the MFF [singular] vectors.  The expression
\req{mff} is in fact used to `generate' singular vectors by applying
analytically continued commutation relations in the universal enveloping
algebra of $\SL2_k$:
\BE
\new\BA{rcl}(J^-_0)^pJ^0_m&=&pJ^-_m(J^-_0)^{p-1}+J^0_m(J^-_0)^p\,,\\ {}
(J^-_0)^pJ^+_m&=&J^+_m(J^-_0)^p+2pJ^0_m(J^-_0)^{p-1}+
p(p-1)J^-_m(J^-_0)^{p-2}\,,
\\{}
J^0_m(J^+_{-1})^p&=&p(J^+_{-1})^{p-1}J^+_{m-1}+(J^+_{-1})^p J^0_m\,,
\\{}
J^-_m(J^+_{-1})^p&=&(J^+_{-1})^p J^-_m + 2p(J^+_{-1})^{p-1}J^0_{m-1}+
p(p-1)(J^+_{-1})^{p-2} J^+_{m-2}-k p\delta_{m-1,0}(J^+_{-1})^{p-1}\,
\EA
\label{properties}\EE
(which are assumed to hold for arbitrary complex $p$).  Upon the repeated use
of these relations all the non-positive-integer powers disappear from the MFF
formula~\req{mff}, and it thus rewrites in a `Verma' form, i.e.\ as an
explicitly Verma module element.

It is only this `Verma' form of $\SL2$ singular vectors, rather than the
`MFF' form~\req{mff}, that can arise from direct computations \cite{[S-sing]}
of topological singular vectors with the help of the Kazama--Suzuki
mapping~\req{GQHks}, \req{Tks}.  However, in view of the actual coincidence
\req{isomorphism} of topological singular vectors with the MFF vectors, an
intriguing question is whether the entire MFF construction would have a
counterpart for the twisted $\N2$ algebra, in the form of a generating
expression for topological singular vectors.

\medskip

Before proceeding to the topological construction, let us note that the
situation described by eqs.~\req{mff}, \req{properties} can be viewed as an
extension of the universal enveloping algebra of $\SL2_k$ by new elements,
{\it denoted\/} as $(J^+_{-1})^p$ and $(J^-_0)^p$, and introducing the
algebraic rules necessary to deal with the new objects; some of these rules
are self-evident due to the `exponential notation'. Note also that not every
commutation relation is defined (for instance, the commutator of
$(J^+_{-1})^p$ with $(J^-_0)^{p'}$ is not, unless $p$ or $p'$ is an integer).
The same idea works for the topological conformal algebra, although more
algebraic rules will have to be written out explicitly.

\section{Topological algebra intertwiners and generalized Verma
modules\label{sec:Inter}}
\subsection{Motivation}\lvm
There are two cases when the `topological' counterpart of the MFF
construction can be built immediately. This happens for a discrete subset of
the  parameter values, namely when the MFF formula~\req{mff} does not require
any manipulations in order to be transformed into a Verma module element; it
can then be rewritten identically in the topological guise, as a formula that
describes a subset of topological singular vectors.  The two cases are
realized either when $s=1$ or for those $r$ and $s$ and {\it integral\/} $t$
for which $|(s-1)t|\leq r$ (so that all exponents in~\req{mff} are positive
integers).  In the first case, the $\ket{S(r,1)}^+$ topological singular
vectors read
\BE\cQ_0\cQ_1\ldots\cQ_{r-1}\,
\cG_{-r}\ldots\cG_{-2}\cG_{-1}\,\ket{\htop^+(r,1)}_{\rm top}
\label{simple}\EE
which under the Kazama--Suzuki mapping becomes just the singular vector
$(J^+_{-1})^r\ket{{\rm MFF}\{r,1\}}\tensor\ket0_{BC}$ (we have thus recovered
a representation that has been known for some time~\cite{[Lerche-pr]}). In
the second case, a generalization of the previous formula is not difficult to
construct, and one finds for the $\nket{T(r,s)}^+$-vector
(see~\req{ST}) the following representation:
\BE\new\BA{rcl}
\ket{T(r,s)}^+&=&\cG_{(\htop-s+1)t-1}
\ldots\cG_{(s-1)t-2}\cG_{(s-1)t-1}\,\cdot\,\\
&&\qquad
\cQ_{-(s-1)t}\ldots\cQ_{(s-\htop-2)t-1}\cQ_{(s-\htop-2)t}\\
&&\qquad\qquad\qquad\ldots\\
&&\qquad\qquad\qquad\cG_{(\htop-1)t-1}\ldots\cG_{t-2}\cG_{t-1}\,\cdot\,\\
&&\qquad\qquad\qquad\qquad
\cQ_{-t}\ldots\cQ_{-\htop t-1}\cQ_{-\htop t}\\
&&\qquad\qquad\qquad\qquad\qquad
\cG_{\htop t-1}\ldots\cG_{-2}\cG_{-1}\,\ket\htop_{\rm top}
\EA\label{N2singheu}\EE
where $\htop=\htop^+(r,s)$ for the appropriate $r$ and $s$; the above
conditions on $t$ and $r$ and $s$ guarantee that the difference between the
mode numbers of the right and the left factors in each group
(`right'${}-{}$`left') is a non-negative integer (or $-1$, in which case the
product evaluates as 1).  This topological singular vector evaluates under
the Kazama-Suzuki mapping precisely as $\ket{{\rm
MFF}\{r,s\}}\tensor\ket0_{BC}$, as is easy to check (each group of factors
in~\req{N2singheu} corresponds to some $(J^\pm_{-1,0})^m$ in~\req{mff}).

\subsection{Generalized Verma modules}\lvm
Taking eq.~\req{N2singheu} as a motivation, one may try to arrive at the
general construction by an `algebraic continuation' of the last formula. One
readily sees that the sought formula should involve a continuation with
respect to {\it modes\/} of the operators.  The topological conformal algebra
commutators can easily be extended to arbitrary (complex) mode numbers.
However, eq.~\req{N2singheu} also tells us that the desired extension would
require objects that represent products of modes of $\cG$ and $\cQ$ with a
{\it non-integral number of factors\/}.

This hint proves to be more constructive than it might  seem, as the desired
objects can in fact be given meaning as intertwining operators
(see footnote~\ref{foot:intertw})
\BE
g(a,b)\quad{\rm and}\quad q(a,b)\quad a,b\in\oC\,.
\label{gq}\EE
They map between {\it generalized Verma modules\/} $\cV_{\theta,\htop}$,
which are related to the ordinary Verma modules by the spectral flow
transform~\cite{[SS],[LVW]} (see also~\cite{[Bea]}).  Namely, the spectral
flow provides an isomorphism between a generalized Verma module
$\cV_{\theta,\htop}$ (to be defined below) and the ordinary Verma module
$V_{\htop-{2\over t}\theta}$ from Sec.~2. At the same time, the topological
conformal algebra is mapped under the spectral flow into an isomorphic
algebra via $\cA_n\mapsto\cA^\theta_n$, where
\BE\new\BA{rcl}
\cL^\theta_n&=&\cL_n+\theta\cH_n+\frac{\ctop}{6}(\theta^2+\theta)
\delta_{n,0}\,, \qquad
\cH^\theta_n=\cH_n+\frac{\ctop}{3}\theta\delta_{n,0}\,,\\
\cQ^\theta_n&=&\cQ_{n-\theta}\,,\qquad\cG^\theta_n=\cG_{n+\theta}\,
\EA\label{U}\EE
(which is a `twisted' form of the corresponding relations from~\cite{[LVW]}).
It is important that the modes of $\cL$ and $\cH$ remain integral after this
transformation. We can then identify the respective $\cL$ and $\cH$
generators in the different modules. The
generalized Verma modules will be considered as modules
over an isomorphic image of the
topological conformal algebra, in which the fermionic generators are realized
as $\cG^\theta_n=\cG_{\theta+n}$, $\cQ^\theta_n=\cQ_{-\theta+n}$, while
$\cL_n$ and $\cH_n$ are `the same' in the different modules:
\BE\new\BA{lclclcl}
\L[\cL_m,\cL_n\R]&=&(m-n)\cL_{m+n}\,,&\qquad&[\cH_m,\cH_n]&=
&\frac{\ctop}{3}m\delta_{m+n,0}\,,\\
\L[\cL_m,\cG_\nu\R]&=&(m-\nu)\cG_{m+\nu}\,,&\qquad&
[\cH_m,\cG_\nu]&=&\cG_{m+\nu}\,,
\\
\L[\cL_m,\cQ_\lambda\R]&=&-\nu\cQ_{m+\lambda}\,,&\qquad&
[\cH_m,\cQ_\lambda]&=&-\cQ_{m+\lambda}\,,\\
\L[\cL_m,\cH_n\R]&=&\multicolumn{5}{l}{-n\cH_{m+n}+\frac{\ctop}{6}(m^2+m)
\delta_{m+n,0}\,,}\\
\L\{\cG_\nu,\cQ_\lambda\R\}&=&\multicolumn{5}{l}{2\cL_{\nu+\lambda}-
2\lambda\cH_{\nu+\lambda}+
\frac{\ctop}{3}(\nu^2+\nu)\delta_{\nu+\lambda,0}\,,}
\EA\qquad
\BA{rcl} m,~n&\in&\oZ\,,\\
\nu&\in&\theta+\oZ\,,\\
\lambda&\in&-\theta+\oZ
\EA
\label{topalgebra1}
\EE
As can be seen, the commutator of, say, $\cG^\theta_n$ and $\cQ^\theta_m$ is
not the same as that of $\cG_n$ and $\cQ_m$. However, the difference can be
eliminated by a redefinition of the $\cL_n$ and $\cH_0$ in any given module.
We will work in what follows with the `common' $\cL_n$ and $\cH_n$
generators.

Then, the highest-weight conditions that determine a generalized
highest-weight state can have one of the two forms:
\BE\new\BA{rclcrcl}
\cL_m&\approx&0\,,\quad m\geq1\,,\quad&
 \cQ_\lambda&\approx&0\,,&\lambda=-\theta+p\,,\quad p=1,2,\ldots\\
\cH_m&\approx&0\,,\quad m\geq1\,,     &
  \cG_\nu&\approx&0\,,&\nu=\theta+p\,,\quad p=0,1,2,\ldots
\EA
\label{ghw}\EE
or
\BE\new\BA{rclcrcl}
\cL_m&\approx&0\,,\quad m\geq1\,,\quad&
 \cQ_\lambda&\approx&0\,,&\lambda=-\theta+p\,,\quad p=0,1,2,\ldots\\
\cH_m&\approx&0\,,\quad m\geq1\,,     &
  \cG_\nu&\approx&0\,,&\nu=\theta+p\,,\quad p=1,2,\ldots
\EA\label{qhw}\EE
However, these actually determine a class of modules (see \cite{[ST4]}) more
general than those that we are going to consider in this paper.  The
generalized {\it topological\/} highest-weight conditions that we will work
with are obtained by strengthening each of the previous ones:
\BE\new\BA{rclcrcl}
\cL_m\ket{\theta,\htop}_{\rm top}&=&0\,,\quad m\geq1\,,\quad&
\cQ_\lambda\ket{\theta,\htop}_{\rm top}&=&0\,,
&\lambda=-\theta+p\,,\quad p=0,1,2,\ldots\\
\cH_m\ket{\theta,\htop}_{\rm top}&=&0\,,\quad m\geq1\,,     &
\cG_\nu\ket{\theta,\htop}_{\rm top}&=&0\,,&\nu=\theta+p\,,\quad p=0,1,2,\ldots
\EA
\label{tophw}\EE
The corresponding states $\nket{\theta,\htop}_{\rm top}$ will be called
the (generalized) topological highest-weight states. Then the generalized Verma
modules $\cV_{\theta,\htop}$ are freely generated from
$\nket{\theta,\htop}_{\rm top}$ by those of the generators from
\req{topalgebra1} that are not declared annihilation operators in
\req{tophw}\,\footnote{We think of the topological central charge
$\ctop\equiv3(t-2)/t$ as fixed; a more rigorous notation for the modules
should be $\cV_{\theta,\htop,t}$.}.  The eigenvalues of the generators
$\cL_0$ and $\cH_0$ from \req{topalgebra1} on a topological highest-weight
state in $\cV_{\theta,\htop}$ are given by
\BE\new\BA{rcl} \cH_0+\theta&\approx&\htop\,\\
\cL_0+\theta\cH_0 +\frac{\ctop}{6}(\theta^2 + \theta)&\approx&0
\EA\label{L0H0}\EE
The ordinary topological Verma modules, where $\cL_0\ket\htop_{\rm top}=0$
and $\cH_0\ket\htop_{\rm top}=\htop\ket\htop_{\rm top}$, are thus a
particular case of a more general situation described by~\req{tophw}
and~\req{L0H0}.

To see which ordinary Verma module is isomorphic to $\cV_{\theta,\htop}$, we
have to use the basis in which the commutation relations of the algebra
become identical to \req{topalgebra}; for $\cH_0$ this amounts to accounting
for the $\theta{\ctop\over3}$ term in \req{U}. We then see that
\BE
\cV^{\phantom{y}}_{\theta,\htop}\sim
V^{\phantom{y}}_{\htop-{2\over t}\theta}\,,
\EE
as claimed above.

\subsection{Algebra of the intertwiners}\lvm
By definition, the mapping of the intertwiners \req{gq} preserves the
highest-weight conditions \req{ghw} {\it or\/} \req{qhw}
when these operators are applied as:
\BE
g(\theta',\theta-1)\,:\,\cV_{\htop,\theta}\rightarrow
\cV_{\htop,\theta'}\,,\qquad
q(-\theta',-\theta-1)\,:\,
\cV_{\htop,\theta}\rightarrow
\cV_{\htop,\theta'}\,,
\label{mappings}\EE
which thus relates the arguments of $g$ and $q$ to the $\theta$-parameters
of the corresponding generalized Verma modules.
We will further
require that the {\it `topological'\/}
highest-weight conditions \req{tophw}
be preserved under the mappings, as this will be needed
in the construction of singular vectors.  As we will see,
when the intertwiners \req{gq} map between the generalized
topological Verma modules, the highest weight states map as
follows. \ $g(\theta',\theta-1)$ and $q(-\theta',-\theta-1)$
take the topological highest-weight state in
$\cV_{\theta,\htop}$ into the topological highest-weight state in
$\cV_{\theta',\htop}$ (i.e., the conditions \req{tophw} are
preserved) if and only if \BE\theta'=\htop t - \theta -
1\qquad{\rm or}\qquad \theta'=(\htop+1)t-\theta-1 \label{iff}\EE
respectively.  A second application of an intertwiner of the
same sort maps back to the original module, while the other
intertwiner maps to a new module:
\begin{equation} \unitlength=1.00mm
\begin{picture}(150,17)
 \put(24.6,0){
    \begin{picture}(150,15)
    \bezier{150}(10,10)(20,14)(30,10)
    \put(28.5,10.85){\vector(3,-2){2}}
    \bezier{150}(10,8)(20,4)(30,8)
    \put(19,13.5){$g$}
    \put(19,3.5){$g$}
    \put(11.9,6.85){\vector(-3,2){2}}
    \end{picture}
  }
 \put(47.3,0){
    \begin{picture}(150,15)
    \put(8,8){$\bullet$}
    \bezier{150}(10,10)(20,14)(30,10)
    \put(28.5,10.85){\vector(3,-2){2}}
    \bezier{150}(10,8)(20,4)(30,8)
    \put(19,13.5){$q$}
    \put(19,3.5){$q$}
    \put(11.9,6.85){\vector(-3,2){2}}
    \end{picture}
  }
 \put(70,0){
    \begin{picture}(150,15)
    \put(8,8){$\bullet$}
    \bezier{150}(10,10)(20,14)(30,10)
    \put(28.5,10.85){\vector(3,-2){2}}
    \bezier{150}(10,8)(20,4)(30,8)
    \put(19,13.5){$g$}
    \put(19,3.5){$g$}
    \put(11.9,6.85){\vector(-3,2){2}}
    \end{picture}
  }
 \put(92.7,0){
    \begin{picture}(150,15)
    \put(8,8){$\bullet$}
    \bezier{150}(10,10)(20,14)(30,10)
    \put(28.5,10.85){\vector(3,-2){2}}
    \bezier{150}(10,8)(20,4)(30,8)
    \put(19,13.5){$q$}
    \put(19,3.5){$q$}
    \put(11.9,6.85){\vector(-3,2){2}}
    \end{picture}
  }
\put(78,3){$\cV_{\theta,\htop}$}
\end{picture}
\label{picture}
\end{equation}
In order to encounter an ordinary Verma module $\cV_{0, \htop}\equiv
V_{\htop}$ among the modules thus obtained from $\cV_{\theta,\htop}$, the
values of $\theta$ should therefore be $\theta_i$, $i=0,1,2,\,\ldots$, where
\BE\new\BA{rcl}
\theta_i=&\left\{
\new\begin{array}{ll}
(\htop-j)t-1\quad&i=2j+1\,,\\
jt\quad&i=2j\,.\\
\end{array}
\right. &
\EA\label{thethetas}\EE

Heuristically, the intertwiners $g(a,b)$ or $q(a,b)$ are a continuation of
the product of modes $\cG_{a}\ldots\cG_{b}$ and $\cQ_{a}\ldots\cQ_{b}$
respectively to arbitrary (complex) $a$ and $b$. This is formalized as the
following property: Whenever the {\it length\/} $\ell=b-a+1$ of an
intertwiner $g(a,b)$ (or $q(a,b)$) becomes a positive integer, that
operator rewrites as the corresponding product of modes
\BE
g(a,a+n)=\prod_{i=0}^n\cG_{a+i}\,,\quad q(a,a+n)=\prod_{i=0}^n\cQ_{a+i}\,,
\quad n=-1,0,1,2,\ldots
\label{integerlength}\EE
(in the case $n=-1$ the product evaluates as 1). As long as the intertwiners
preserve the topological highest-weight conditions, the intertwiners of
positive integral length, once they send the highest-weight state into a
state {\it in the same\/} module, are in fact singular vectors.  On the other
hand, $g(a,a-n)$ for a positive integral $n$ does {\it not\/} evaluate as 1
unless $n=1$.  The intertwiners $g(a,a-n)$ of negative integral length $-n+1$
represent {\it co\/}singular vectors.

The intertwiners inherit a number of further properties from the products of
modes. These properties will be imposed, in addition to \req{integerlength},
as a consistent set of commutation relations and certain rearrangement rules
for $q(a,b)$ and $g(a,b)$ with arbitrary complex $a$ and $b$.  We will
formulate these properties for the $g$ intertwiners, the `$q$'-case being
completely analogous.  To begin with, these are the relations
\BE
g(a,b-1)\,g(b,c)=g(a,c)\,,\qquad g(a+1,a)=1\,.
\label{Un2g}\EE
It follows then that, in particular,
\BE
\cG_a\,g(a+1,b) = g(a,b)\,,\quad g(a,b)\,\cG_{b+1} = g(a,b+1)\,.
\label{Glue}\EE
Further,
\BE
\cG_a\,g(b,c)=0\,,\qquad a-b=0,1,2,\ldots\,,\quad c-b+1\neq0,1,2,\ldots\,,
\label{leftkill}
\EE
and, similarly,
\BE
g(a,b)\,\cG_c = 0\,,\qquad b-c=0,1,2\ldots\,,\quad b-a+1\neq0,1,2,\ldots\,.
\label{rightkill}
\EE
(in the case of a positive integral length $c-b+1$ or $b-a+1$, the reduction
\req{integerlength} has to performed first, after which the vanishing
conditions become the usual ones in the universal enveloping of the $\N2$
algebra).

Relations similar to \req{Un2g}--\req{rightkill} hold for the $q$
intertwiners (provided $\cG$ is replaced with $\cQ$).  Eqs.~\req{leftkill},
\req{rightkill} can be viewed as a manifestation of the Pauli principle,
namely that $g(a,b)$ and $q(a,b)$ represent states from $a$ to $b$ filled
with the corresponding fermions.  The relations \req{Un2g} and \req{Glue}, as
well as those that we are going to introduce, can be checked to hold when the
intertwiners involved reduce to a product of the corresponding modes (it
should be clear, however, that the inverse is far from being true: not every
relation that holds in the universal enveloping of~\req{topalgebra} extends
to the intertwiners!).

To see how the `Pauli principle' allows us to arrive at the necessary
commutation relations, consider for instance commuting the mode $\cH_p$, with
$p$ an integer $\geq1$, through $g(a,b)$. In the universal enveloping
algebra, with $a$ and $b$ positive integers such that $b>a$, we have from
\req{topalgebra}
\BE
[\cH_p,\,g(a,b)]\equiv[\cH_p,\,\prod_{i=a}^b\cG_i]=
\sum_{j=a}^b\Bigl(\prod_{i=a}^{j-1}\cG_i\Bigr)
\,\cG_{j+p}\,
\Bigl(\prod_{i=j+1}^b\cG_i\Bigr)
\label{hint1}\EE
However, the terms in the sum vanish whenever $j+p$ (which so far is an
integer) is inside the segment $[j+1, b]$. Therefore, the RHS of the last
formula contains only the terms with $j=b-p+1,\ldots,b$, hence
\BE
\cH_p\,g(a,b) = g(a,b)\,\cH_p +
\sum_{j=1}^pg(a,b-p+j-1)\,\cG_{b+j}
\prod_{i=1}^{p-j}\cG_{b-p+j+i}\,,\qquad p\geq1\,.
\label{HRight}\EE
When written in this form, the formula can be used for {\it arbitrary\/} $a$
and $b$, since the RHS contains an integral number of terms and the product
involves an integral number of factors, independently of the values of $a$
and $b$. It is in this form that we fix the commutation relation
$[\cH_p,\,g(a,b)]$.  For $\cH_0$, the definition is motivated by even simpler
relations
\BE
\bigr[\cH_0,\,\prod_{i=m}^n\cG_i\bigr]=
(n-m+1)\,\prod_{i=m}^n\cG_i\,,\qquad
\bigr[\cH_0,\,\prod_{i=m}^n\cQ_i\bigr]=
(-n+m-1)\,\prod_{i=m}^n\cQ_i
\label{H0}\EE
which in an obvious way rewrite in terms of $g(m,n)$ and
$q(m,n)$  and then extend to arbitrary $m$ and~$n$.  The formulae that
describe commutators of $\cL_{\geq0}$ with $g(a,b)$ (and $q(a,b)$), can be
derived similarly to \req{HRight}, with use being made of the commutators
$[\cL_n,\,\cG_m]$ and $[\cL_n,\,\cQ_m]$ from~\req{topalgebra}:
\BE
[\cL_0,\,g(a,b)]=-\half(a+b)(b-a+1)\,g(a,b)\,,\qquad
[\cL_0,\,q(a,b)]=-\half(a+b)(b-a+1)\,q(a,b)\,
\EE
and
\BE
\cL_p g(a,b)=g(a,b)\cL_p+\sum_{l=0}^{p-1}g(a,b-l-1)
\left[\cL_p\,,\,\cG_{b-l}\right]\,\cG_{b-l+1}\ldots \cG_b\,,\quad p\geq1\,.
\label{LRight}\EE

Finally, when checking the highest-weight conditions, one also uses the
following consequence of the Pauli principle:
\BE
\cG_{\theta+n}\,g(\theta,\theta'-1)\,
q(-\theta',-\theta''-1)\,\ldots\,g(\tilde\theta,-1)\,
\ket\htop_{\rm top}=0\,,\qquad n=1,2,\ldots
\label{Qq}\EE

In addition, one can evaluate
$\cQ_{-\theta}\,g(\theta,\theta'-1)\,q(-\theta',-\theta''-1)\,
\ldots\,g(\tilde\theta,-1)\,\ket\htop_{\rm top}$ by writing the
leftmost intertwiner as $\cG_{\theta}\,g(\theta+1,\theta'-1)$, whence
\BE\new\BA{l}
\cQ_{-\theta}\,g(\theta,\theta'-1)\,q(-\theta',-\theta''-1)\,
\ldots\,g(\tilde\theta,-1)\,\ket\htop_{\rm top}={}\\
\qquad\{\cQ_{-\theta},\,\cG_{\theta}\}\,
g(\theta+1,\theta'-1)
\,q(-\theta',-\theta''-1)\,\ldots\,g(\tilde\theta,-1)\,\ket\htop_{\rm top}
\EA\label{cross}\EE
which holds provided $\theta$, $\theta'$, $\ldots$, $\tilde\theta$ match as
in \req{thethetas}.  Here, the commutator $\{\cQ_{-\theta},\,\cG_{\theta}\}$
should be taken from \req{topalgebra1} and the resulting modes of $\cL$ and
$\cH$ then evaluated as explained above (`symmetric' formulae hold for
$q\leftrightarrow g$). This then allows us to check whether or not the
topological highest-weight conditions \req{tophw} are preserved under the
action of $g$ and $q$, and this is how eqs.~\req{iff} have been obtained.

Further, using the equations \req{HRight} and \req{LRight}, we conclude from
\req{cross} that
\BE\cQ_{-\theta+n}\,g(\theta,\theta'-1)\,q(-\theta',-\theta''-1)\,
\ldots\,g(\tilde\theta,-1)\,\ket\htop_{\rm top}=0\,,\quad n=1,2\ldots
\label{inner}\EE
which may be interpreted again in the spirit of the Pauli principle: all the
`inner' commutators in \req{inner} obtained while plugging $\cQ_{-\theta+n}$
on the right will necessarily produce terms that vanish either due to the
previously established highest-weight conditions, or due to the Pauli
principle.

\medskip

For negative-moded $\cH$ and $\cL$, the following formulae are derived by a
simple modification of the argument~\req{hint1}--\req{HRight}:
\BE
\new\BA{rcl}
g(a,b)\cL_p&=&\cL_p g(a,b)+\sum_{l=0}^{-p-1}\cG_a\ldots \cG_{a+l-1}
\left[\cG_{a+l}\,,\,\cL_p\right]\,g(a+l+1,b)\,,\quad p\leq-1\,\\
g(a,b)\cH_p&=&\cH_p g(a,b)+\sum_{l=0}^{-p-1}\cG_a\ldots \cG_{a+l-1}
\left[\cG_{a+l}\,,\,\cH_p\right]\,g(a+l+1,b)\,,\quad p\leq-1\,.
\EA\label{LHLeftg}
\EE
These relations will in fact be the basic tool in what follows (their
analogues with $g$ replaced by $q$ are straightforward to formulate). As we
will see, eqs.~\req{LHLeftg} are a `topological' counterpart of the $\SL2$
formulae~\req{properties}. However, while eqs.~\req{properties} allow one to
commute the powers of $J^+_{-1}$ and $J^-_0$ through any of the generators
$J^{\pm,0}_m$, in the topological case we will never need the analogue
of~\req{LHLeftg} with $\cL_p$, $p=-1,-2,\ldots$, replaced with
$\cQ_{-\theta+p}$ or $\cG_{\theta+p}$. The mechanism that spares us commuting
the fermions through the intertwiners (the appearance of an
`$\cL$-$\cH$-skeleton') will be explained below.

\section{Topological singular vectors and generalized Verma
modules\label{sec:Gen}}\lvm
The question of whether a module admits a singular vector can be reformulated
in terms of the existence of an intertwiner from that module to itself.  The
mappings implemented by the intertwiners have been tuned (see
\req{thethetas}) to preserve the topological highest-weight conditions
\req{tophw}. Thus, moving along the arrows in the diagram \req{picture}, we
are interested in the case when a sequence of oscillating $q$- and $g$-
arrows makes a loop:
\begin{equation}
\begin{picture}(150,80)
\put(79.0,70.0){$\bullet$}
\put(78.0,77.0){$\cV_{\theta,\htop}$}
\put(85.0,71.5){\vector(4,-1){30}}
\put(95.0,62.0){$g$}
\put(115.5,60.0){$\bullet$}
\put(122.0,60.0){\vector(2,-1){30}}
\put(130.5,48.0){$q$}
\put(152.0,40.0){$\bullet$}
\put(157.0,38.0){\vector(1,-2){12}}
\put(168.0,0.0){$\vdots$}
\put(63.0,62.0){$g$}
\put(41.5,60.0){$\bullet$}
\put(48.0,64.5){\vector(4,1){30}}
\put(20.5,45.0){$q$}
\put(5.0,40.0){$\bullet$}
\put(11.0,45.0){\vector(2,1){30}}
\put(-6.0,15.0){\vector(1,2){12}}
\put(-7.0,0.0){$\vdots$}
\end{picture}
\label{ring}
\end{equation}
Then, $\cV_{\theta,\htop}$ will have a singular vector.  In the following, we
will apply this strategy to constructing the topological singular vectors in
the usual Verma modules, which will mean taking $\cV_{\theta,\htop}$ in
\req{ring} to be $\cV_{0, \htop}\equiv V_{\htop}$, the module built over
$\nket{\htop}_{\rm top}$.  Thus, after $2\ell+1$ steps along the arrows in
\req{ring}, we arrive at a state (see~\req{thethetas})
\begin{eqnarray}
&&\ket{\htop, t, \ell}_{\rm top}^+={}\nonumber\\
&&\qquad g((\htop-\ell)t-1,\ell t-1)\,
q(-\ell t,(\ell -1-\htop)t)~\ldots\label{N2sing0}\\
&&\qquad\qquad g((\htop-1)t-1,t-1)\, q(-t,-\htop t)\,
g(\htop t-1,-1)\,\ket{\htop}_{\rm top}\nonumber
\end{eqnarray}
The condition for this to belong to $V_{\htop}$ and be non-vanishing is that
$(\htop-\ell)t-1$ be a negative integer, say $-r$.  We thus recover the
parametrization for $\htop=\htop^+$ as in~\req{htoprs} in terms of two
positive integers $r$ and $s=\ell+1$ (the formula for $\htop^-$ is recovered
similarly, by considering a loop starting and ending with a $q$ intertwiner).
{}From now on, we will denote $V_{\htop^\pm(r,s)}$ as $V^\pm_{r,s}$.

By writing the state \req{N2sing0} for a closed loop, we arrive at the
formula for topological singular vectors in the topological Verma
module~$V_{r,s}^+$:
\begin{eqnarray}
\ket{T(r,s)}^+&=& g(-r,(s-1)t-1)\,
q(-(s-1)t,r-1-t)\,\ldots{}\label{N2sing}\\
{}&{}&\quad{}g((s-2)t-r,t-1)\,
q(-t,r-1-t(s-1))\, g((s-1)t-r,-1)\,\ket{\htop^+(r,s)}_{\rm top}\,,\nonumber\\
&&r,s\in\oN\nonumber
\end{eqnarray}
Every `intermediate' module appears in~\req{ring} twice, which
guarantees that we end up with a singular vector in the original
module\footnote{and also that `intermediate' singular vectors
$\nket{T(r,s'<s)}^\pm$ follow from~\req{N2sing}, as we will see in
section~\ref{sec:Rec}.}: denoting the intermediate modules as
$\cV^\pm_{r,s}$, we have
\BE\BA{l}
V^+_{r,s}\supset\cV^{-,\theta_{2s-1}}_{r,s}\leftarrow
\cV^{+,\theta_{2s-2}}_{r,s-1}\ldots\leftarrow
\cV^{\mp,\theta_s}_{r,1}\leftarrow\cV^{\pm,\theta_{s-1}}_{r,1}\leftarrow
\ldots\leftarrow\cV^{-,\theta_1}_{r,s-1}\leftarrow V^+_{r,s}\,,\\
\phantom{{V^+_{r,s}\supset\cV^{-,\theta_{2s-1}}_{r,s}\leftarrow
\cV^{+,\theta_{2s-2}}_{r,s-1}\ldots\leftarrow{}}}
\,\cap\\
\phantom{{V^+_{r,s}\supset\cV^{-,\theta_{2s-1}}_{r,s}\leftarrow
\cV^{+,\theta_{2s-2}}_{r,s-1}\ldots\leftarrow{}}}
\cV^{\pm,\theta_{s-1}}_{r,1}
\EA\label{longchain}
\EE
(with $\cV^{\mp,\theta_i}_{r,s}$ standing for the image of $\cV^{\mp}_{r,s}$
under the spectral flow transformation with the $\theta_i$ parameter taken
from~\req{thethetas}).  In the `centre' of \req{longchain}, we have the
mapping given by the intertwiner $g(a,b)$ (for $s$ odd) or $q(a,b)$ (for $s$
even) of positive integral length $b-a+1=r$.  Therefore, while moving from
the right-hand end to the centre of~\req{longchain}, the application of the
intertwining operators decreases the value of $s$ and leads to a module in
which $s=1$ and hence the singular vectors have the simplest
form~\req{simple}. It is this $\nket{T(r,1)}$ vector in the appropriate
generalized Verma module, given by the product of $r$ modes of $\cG$ (or
$\cQ$), that we encounter in the `centre' of the formula~\req{N2sing}. The
left `half' of the diagram \req{longchain} does bring us back into the
original module.

Note also that the topological $U(1)$ charge (the eigenvalue of $\cH_0$ on the
generalized highest-weight state) is mapped under~\req{longchain}
as
\BE
\ldots\leftarrow h-4\leftarrow\frac{2}{t}-h+2\leftarrow h-2
\leftarrow\frac{2}{t}-h\leftarrow h
\label{hmap}\EE
which suggests an underlying structure similar to the affine Weyl group
one has in the $\SL2$ case.

Analogous formulae, with $g\leftrightarrow q$, and $\cQ\leftrightarrow\cG$
and ${}+\leftrightarrow-{}$, exists for $\ket{T(r,s)}^-$: the counterpart of
\req{N2sing0} reads
\begin{eqnarray}
&&\qquad
q(-(\htop+\ell+1)t + 1, \ell t -1)\,g(-\ell t, (\htop + \ell)t - 2)\,~\ldots\\
&&\qquad\qquad q(1 - 2 t - \htop t, -1 + t)\,g(-t, -2 + t + \htop t)\,
q(1 - t - \htop t, -1)\,\ket{\htop}_{\rm top}\,,\nonumber
\end{eqnarray}
whence
\begin{eqnarray}
\ket{T(r,s)}^-&=&
q(-r, -1 - t + s t)\,g(t - s t, -1 + r - t)\,\ldots\label{Tminus}\\{}&{}&{}
\qquad q(-r - 2 t + s t, -1 + t) \, g(-t, -1 + r + t - s t)\,
q(-r - t + s t, -1)\,\ket{\htop^-(r,s)}_{\rm top}\nonumber
\end{eqnarray}

The action of the intertwiners does preserve the topological highest-weight
conditions \req{tophw} at every step, therefore these conditions hold for the
states \req{N2sing} and~\req{Tminus}.  In the next section we will describe
an operational scheme allowing us to rewrite any state of the
form~\req{N2sing} and~\req{Tminus} as a Verma module element, i.e.\ as a
polynomial in $\cL_{m<0}$, $\cH_{m<0}$, $\cG_{m<0}$ and $\cQ_{m<0}$ ($m$
integral) acting on $\nket{\htop^\pm(r,s)}_{\rm top}$, with no intertwiners
involved. This will amount to a rearrangement of~\req{N2sing}, \req{Tminus}
by a systematic use of the properties of the intertwiners from
section~\ref{sec:Inter}.

\section{Back to the Verma modules\label{sec:Back}}\lvm
We have seen that the general construction for topological singular vectors
satisfies the required highest-weight conditions. In this section, we will
transform \req{N2sing} into an explicitly `Verma' form. (The other half of
the topological singular vectors, eqs.~\req{Tminus}, are treated similarly).

Let us consider for definiteness the case when the $r$ modes in the centre of
\req{N2sing} are $\cQ$ modes:
$q(-{st\over2},-{st\over2}+r-1)=\cQ_{-{st\over2}}\ldots\cQ_{-{st\over2}+r-1}$.
As follows from~\req{thethetas} and~\req{N2sing}, the lengths of the
intertwiners $g(a,b)\cdot\ldots\cdot g(c,d)$ that occupy symmetric positions
with respect to the centre of the formula~\req{N2sing} always add up to a
positive integer, namely to $2r$:  $(d-c+1)+(b-a+1)=2r$ (the same is true for
any `symmetric' pair $q(a,b)\ldots q(c,d)$ as well).
The aim of the manipulations that follow is to make these intertwiners meet
in such a way that the formula \req{Un2g} would apply, which would then allow
us to replace the resulting intertwiner of integral length with a product of
modes.

By repeatedly (precisely $r$ times) applying to $g(a,b)$ the identical
transformation from eq.~\req{Glue},
\BE
g(a,b)\to g(a,b-1)\cG_b\,,\label{identical}\EE
we bring it to the form of \ $g(a, c-1)$ times a product of modes of $\cG$.
Thus, in the case we have chosen for definiteness, the centre of
the formula~\req{N2sing} rewrites as
\BE\new\BA{l}
g(-r+\frac{st}{2}-t,\frac{st}{2}-1)\,q(-\frac{st}{2},-\frac{st}{2}+r-1)\,
g(\frac{st}{2}-r,\frac{st}{2}-t-1){}\\
{}=g(\frac{st}{2}-t-r,\frac{st}{2}-r-1)\,\\
\qquad{}\cdot\underbrace{\cG_{{st\over2}-r}\,\ldots\,\cG_{{st\over2}-1}}\,
\cQ_{-{st\over2}}\,\ldots\,\cQ_{-{st\over2}+r-1}
g(\frac{st}{2}-r,\frac{st}{2}-t-1)
\EA\EE

According to \req{tophw}, the underbraced operators are annihilators with
respect to the generalized highest-weight state
$g(\frac{st}{2}-r,\frac{st}{2}-t-1)\ldots \nket{\htop^+(r,s)}_{\rm top}$ that
we have on the right of the modes.  This is a general fact, as can be seen
from the structure of~\req{N2sing} and the `highest-weight'
conditions~\req{tophw}: rewriting a `symmetric' pair of intertwiners
$g(a,b)\,\ldots\,g(c,d)$ as \ $g(a,c-1)\bigl(\prod\cG\bigr)\,\ldots\,g(c,d)$,
we always have in $\prod\cG$ precisely $r$ modes that annihilate the state
$g(c,d)\ldots\nket{\htop}$. All these annihilators have to be commuted to the
right and killed on the generalized highest-weight state.  Generically, more
annihilation operators would appear from the commutators, and these, too,
have to be moved right.  Any $\cL_{\geq0}$ or $\cH_{\geq0}$ that appears in
the course of commutation should also be evaluated on the vacuum, according
to \req{tophw} and~\req{L0H0}.

The next step is to observe that all the operators $\cG$ that might be
present between the intertwiners after all the highest-weight conditions have
been used, are such that the rule~\req{leftkill} applies. This is again true
in general, because the modes in question have been split off from the
left-hand intertwiner, and then commuted through the other operators which
necessarily were creators, hence the mode numbers could only decrease and
therefore the condition required in \req{leftkill} fulfills.

Therefore, in the `$\cG$'-case under consideration, we can apply the
rules~\req{leftkill} after having commuted all the $\cG$'s to the left.
Practically, it suffices to commute them as far left as (and through) the
leftmost $\cQ$ operator.  We will then be left with a polynomial in modes of
{\it only\/} $\cL$ and $\cH$.

This is a crucial circumstance because the $\cL$ and $\cH$ modes can be
commuted through the left intertwiner according to the
formulae~\req{LHLeftg}. When this is done, there will be no more modes
between the two $g$ intertwiners, and the formula~\req{Un2g} will apply.  The
resulting intertwiner $g(\mu,\nu)$  will {\it always\/} have an integral
length $\lambda=\nu - \mu +1$, as can be seen from~\req{N2sing}.  Whenever
$\lambda$ is non-negative, the intertwiner gets replaced by a product of
modes,
\BE g(\mu,\nu)=\prod_{i=0}^{\lambda-1}\,\cG_{\mu+i}\,,\quad{\rm iff}\quad
\lambda\equiv\nu - \mu +1= 0, 1, 2,\ldots
\label{Un1g}\EE
and the innermost $g$ intertwiners have therefore been completely eliminated,
resulting in a sum of terms that contain only the modes of the topological
algebra generators. In each of these terms, one can now apply the same
strategy to the embracing $q$-intertwiners, and so on by induction.  However,
it follows from \req{LHLeftg} that when $\cL_n$ or $\cH_n$ is carried through
$g(a,b)$, it gives rise to a $g$ intertwiner of a smaller length, and this is
the origin of negative-length intertwiners.  In some terms we would thus have
$g(\mu,\nu)$ of a negative integral length instead of~\req{Un1g}.  It thus
remains to give meaning to the remaining intertwiners of negative integral
length.

Here we can be guided by associativity in the universal enveloping algebra
and use it together with the generalized highest-weight and eigenvalue
conditions given above.  Then the negative-length intertwiners are `resolved'
as follows.  Consider the state
\BE\new\BA{l}
g(\mu-1,\theta_{n-1} - 1)\,q(-\theta_{n-1}, -\theta_{n-2}-1)\,\ldots\,
g(\theta_1,-1)\,\ket{\htop}_{\rm top}=\\
\qquad\cG_{\mu-1}g(\mu,\theta_{n-1} - 1)
\,q(-\theta_{n-1}, -\theta_{n-2}-1)\,\ldots\,
g(\theta_1,-1)\,\ket{\htop}_{\rm top}
\EA\EE
and evaluate on it the operator $\cQ_{-\mu+1}$ by first commuting it with
$\cG_{\mu-1}$ as \ $\{\cQ_{-\mu+1},\,\cG_{\mu-1}\}={}\\ {\ctop\over3}(-1 +
\mu)\mu -2(1 - \mu)\cH_0 + 2\cL_0$ and then exploiting the highest-weight
conditions~\req{L0H0}. We thus arrive at the relation
\BE
g(\mu,\theta_{n-1} - 1)\ldots\,g(\theta_1,-1)\,\ket{\htop}_{\rm top}=
{1\over\Lambda(\mu,\theta_{n-1},\htop)}\,
\cQ_{1 - \mu}\,g(\mu - 1, \theta_{n-1} - 1)\ldots\,
g(\theta_1,-1)\,\ket{\htop}_{\rm top}
\label{Fill}\EE
where
$\Lambda$ is the eigenvalue of ${\ctop\over3}(-1 + \mu)\mu -2(1 - \mu)\cH_0 +
2\cL_0$ on the LHS of~\req{Fill},
\BE
\Lambda(\theta+N,\theta,\htop)=(N-1)\left(2\frac{\ctop}{3}\theta+
\frac{\ctop}{3}N+2\htop-2\theta\right)\,.
\label{Lambda}\EE
The result of applying \req{Fill} to intertwiners of negative length is that
the length increases by one, $g(\mu, *)\to g(\mu-1, *)$.  The
rearrangement~\req{Fill} therefore applies successively to the states
\BE
({\sl modes\/})\,g(\theta_{n-1} + N,\theta_{n-1} - 1)\,
q(-\theta_{n-1}, -\theta_{n-2}-1)\,\ldots\,
g(\theta_1,-1)\,\ket{\htop}_{\rm top}
\EE
starting with the higher $N$, down to $\N2$. The $\cG$-operators that can be
present among the {\sl modes\/}, are of the form $\cG_{\theta_{n-1}+p}$ with
integral $p$. If, in particular, $\cG_{\theta_{n-1}+N-2}$ is encountered,
then the rule~\req{Glue} should be applied, which would increase the
length of the remaining intertwiner even further.

The rule \req{Fill} is thus used repeatedly, allowing us to evaluate
$g(\theta_{n-1} + m,\theta_{n-1} - 1)\ldots\,\nket{\htop}_{\rm top}$ in terms
of $g(\theta_{n-1} + m - 1,\theta_{n-1} - 1)\ldots\,\nket{\htop}_{\rm top}$
for $m\geq2$.  A crucial fact is that after all these rearrangements have
been applied in the course of evaluation of a singular vector, {\it all\/}
the remaining states
\BE
({\sl modes\/})\cdot g(\theta_{n-1} + 1,\theta_{n-1} - 1)\,
q(-\theta_{n-1}, -\theta_{n-2}-1)\,\ldots
g(\theta_1,-1)\,\ket{\htop}_{\rm top}
\EE
cancel! This cancellation is a non-trivial property of the whole scheme; a
related fact is that using the formula \req{Fill} always preserves the
coefficients in $t^{-M}\cdot\oQ[t]$ (polynomials with rational coefficients
times a certain negative power of $t$ that may come out of the
highest-weights \req{htoprs}), i.e.\ no rational dependence on $t$ arises,
apart from a possible $t^{-M}$.

The rule~\req{Fill} gives the prescription to deal with intertwiners of
negative integral lengths when evaluating the singular vectors.  We see that a
$g$ intertwiner of a negative integral length $L$, which heuristically stands
for the `absence of ($-L$) \ $\cG$-modes', does basically reduce on a
generalized highest weight state to an insertion of a product of modes of the
other fermion, $\cQ$, times an overall normalization (more terms come out of
the algebra \req{topalgebra1} when using \req{Fill} repeatedly). Such a
rearrangement would not work for arbitrary expressions with negative-length
intertwiners, since the latter do in general take us outside a given Verma
module.  Consider, for example, $\nket{X}=g(1,-1)\nket{\htop}_{\rm top}$,
which is formally a cosingular vector with respect to $\nket{\htop}_{\rm
top}$:  $\nket{\htop}_{\rm top}=\cG_0\nket{X}$, Obviously, it would
generically not be in the Verma module built upon $\nket{\htop}_{\rm top}$.

\smallskip

We see that the pair of intertwiners $g(a,b)\cdot({\rm modes\/})\cdot g(c,d)$
in \req{N2sing} or \req{Tminus} do always `decay' into modes of topological
conformal algebra generators, and thus the formula for topological singular
vector rewrites as a sum of several terms, in each of which the number of
intertwiners has decreased by two.

\medskip

The procedure of `eliminating' the intertwining operators symmetric with
respect to the centre of \req{N2sing} then repeats mutatis mutandis (with
$q\leftrightarrow g$, $\cQ\leftrightarrow\cG$, etc.). The rules formulated
above (together with their `$q$-mirror') are sufficient to carry the
evaluation to the end.  Eventually, we will be left with a singular vector in
the {\it Verma\/} module.  The expressions obtained in the course of this
evaluation acquire more and more complicated structure as a result of
commuting the fermionic operators $\cQ$ or $\cG$ first to the generalized
highest-weight state and then, for the remaining modes, to the intertwiner
standing on the left of the modes.

Therefore, the expression~\req{N2sing} has been given the meaning of a
general algebraic construction for topological singular vectors in Verma
modules.

\section{Topological singular vectors via recursion relations\label{sec:Rec}}
\lvm
As noted above, a very useful reformulation of the procedure described in the
last section can be given in terms of recursion relations that allow one to
construct a singular vector $\ket{S(r,s)}^\pm$ out of $\ket{S(r,s-1)}^\mp$.

Observe that the above step-by-step evaluation of singular vector
$\ket{S(r,s)}^+$ according to eq.~\req{N2sing} has a matrioshka structure:
`inside' the vector $\ket{S(r,s)}^+$ there sit `smaller' singular vectors
$\ket{S(r,s')}^\pm$ with $1\leq s'\leq s$ and $+$ or $-$ oscillating from one
step to another; in the centre, as noted in the previous section, there is
the $r1$ singular vector.  The `Verma' forms of these `intermediate' singular
vectors are arrived at as soon as a given pair of intertwiners $g(a,
b)\,\ldots\,g(c,d)$ or $q(a, b)\,\ldots\,q(c,d)$ is `eliminated' (one
actually applies also the (inverse) spectral flow transform~\req{U} to the
modes of $\cL$, $\cH$, $\cG$ and $\cQ$,  sandwiched between the remaining
intertwiners).

Since each `intermediate' expression is therefore a singular vector, we can
simply take an {\it arbitrary\/} singular vector (not necessarily an $r1$) as
a starting point for producing the higher ones (those with bigger~$s$).  Then
the scheme we applied in Section~\ref{sec:Back} reformulates as follows.

Assume for definiteness that we start with a `$-$'-vector $\nket{S(r,s-1)}^-$
in the Verma module $V^-_{r,s-1}$ over $\nket{\htop}_{\rm
top}\equiv\nket{\htop^-(r, s-1)}_{\rm top}$; let us then construct
$\nket{S(r,s)}^+$ out of it.  Taking $\nket{S(r,s-1)}^-$, it will be assumed
that it is written in the form when all the annihilation operators have been
evaluated on the highest-weight state, therefore the singular vector is given
by a polynomial in creation operators acting on the appropriate
highest-weight state.  Then the following sequence of steps applies (two
examples are given in the Appendix).

\medskip

{\sc Step~1}. Apply to $\nket{S(r,s-1)}^-$ the spectral flow
transformation
\BE\new\BA{rclcrcl}
\cQ_n &\mapsto& \cQ_{n - \theta}\,,&&\cL_n &\mapsto& \cL_n + \theta\cH_n\,,\\
\cG_n &\mapsto& \cG_{n + \theta}\,,&&\cH_n &\mapsto& \cH_n\,,
\EA\label{spectral}\EE
with
\BE
\theta = (s-1)t - r
\EE
(in the case $\nket{S(r,s-1)}^+\to \nket{S(r,s)}^-$ we would have $\theta =
-(s-1)t + r$ instead). Note that since there have been only the creation
operators in the expression for $\nket{S(r,s-1)}^-$, we would not encounter
$\cL_0$ and $\cH_0$, therefore $c$-number terms from~\req{U} will not appear.
The highest weight state has to be transformed accordingly:
\BE
\ket{\htop}_{\rm top} \to g(\theta,-1)\,\ket{\frac{2}{t} - \htop}_{\rm top}\,,
\EE
with the above $\theta$; in the case $\nket{S(r,s-1)}^+\to \nket{S(r,s)}^-$
this would be $\ket{\htop}_{\rm top} \to q(-\theta,-1)\ket{\frac{2}{t} - 2 -
\htop}_{\rm top}$. In this way, we obtain a singular vector in the
generalized Verma module $\cV^-_{r,s-1}$.

\smallskip

{\sc Step~2}. To return to the module $V^+_{r,s}$ where we want to find a
singular vector, apply to the expression resulting from Step~1 the
intertwiner $g(-r, \theta - 1)$ (respectively $q(-r, -\theta - 1)$).  In the
resulting expression $g(-r, \theta - 1)\,({\sl
modes\/})\,g(\theta,-1)\,\ket{\htop^+(r,s)}_{\rm top}$, we use relations
\req{leftkill} by commuting the modes of $\cG$ on the left.  This will
eliminate {\it all\/} $\cG$- and $\cQ$- modes, leaving us with a polynomial
in negative integral modes of $\cL$ and $\cH$ inside $g(-r, \theta -
1)\,\ldots\,g(\theta,-1)\,\ket{\htop^+(r,s)}_{\rm top}$.  This polynomial
will be referred to as a `skeleton' of the singular vector $\nket{S(r,s)}^+$.

\smallskip

{\sc Step~3}. It remains to use relations \req{LHLeftg} to commute the modes
of $\cL$ and $\cH$ through the intertwiner $g(-r, \theta - 1)$ so as to make
the two intertwiners meet.  To \ $g(-r + p + 1, \theta-1)\,g(\theta,-1)=g(-r
+ p + 1, -1)$ we apply~\req{Un1g} whenever possible, that is when $p + 1 \leq
r$ (recall that both $p$ and $r$ are positive integers!). As a result, we
will be left with two kinds of terms: \ i)~those which contain no
intertwiners,
\BE
\cP(\cL, \cH)\,
\cG_{n_1}\,\ldots\,\cG_{n_r}
\,\ket{\htop^+(r,s)}_{\rm top}\,,
\EE
and are therefore in the Verma module $V^+_{r,s}$ \ ($\cP$ is a polynomial in
the negative modes and $n_i$ are negative integers), and \ ii)~those
involving intertwiners of {\it negative\/} integral length:
\BE \cP(\cL, \cH)\,
\cG_{n_1}\,\ldots\,\cG_{n_{r+N}}
\,g(N,-1)\,\ket{\htop^+(r,s)}_{\rm top}\,,
\label{weird}\EE
where $N=1, 2,\ldots, r(s-2)$.
Starting with higher $N$, we then replace
\BE
g(N, -1)\,\ket{\htop^+(r,s)}_{\rm top}= \,
{t\over2(N - 1)(\htop^+(r,s) t - N)}\,
\cQ_{1 - N}\,g(N - 1, -1)\,\ket{\htop^+(r,s)}_{\rm top}\,,\quad N\geq2
\label{Fill1}\EE
(which is a specification of \req{Fill}, \req{Lambda}) and commute the product
$(\prod\cG)$ in \req{weird} to the right~\footnote{Note that the
normalization factor in \req{Fill1} rewrites as
$$
{t\over2(N - 1)(1 - N - r + (s - 1)t)}\,,
$$
in which form it is equally applicable in a similar derivation of
$\nket{T(r,s)}^-$ from $\nket{S(r,s-1)}^+$. Clearly, as was the case with
formula~\req{Fill}, the use of~\req{Fill1} does never give rise to a
rational dependence on $t$, since the terms to which~\req{Fill1} is applied
are proportional to the respective factor $(1 - N - r + (s - 1)t)$.}. Then,
in some terms the combination $(\prod\cG)\,g(N-1, -1)$ would allow us to
apply \req{Glue}, which would give one of the $g(N-2,-1)$, $g(N-3,-1)$,
$\ldots$, $g(0,-1)$ intertwiners.
In the latter case ($g(0,-1)=1$) the intertwiner will
have disappeared. In the terms that would still contain $g(N-1, -1)$,
$\ldots$, $g(2, -1)$, we apply \req{Fill1} and the corresponding
rearrangements again, until we end up with having, on top of a state from the
Verma module $V^+_{r,s}$, only the terms that contain $g(1, -1)$. However,
{\it all the latter cancel against the different terms\/}, and we are
therefore left with a state in $V^+_{r,s}$.

\smallskip

This state is the
singular vector $\nket{T(r,s)}^+$, from which the `conventional' singular
vector follows as $\nket{S(r,s)}^+=
\cQ_0\ldots\cQ_{r-1}\cdot\nket{T(r,s)}^+$.

\smallskip

The $\nket{T(r,s)}^-$ vector is derived similarly from $\nket{S(r,s-1)}^+$.

\medskip

To make the construction of this section complete (sufficient for finding any
topological singular vector), we have to quote the simplest $T$-vectors
$\nket{T(r,1)}^\pm$:
\BE\new
\BA{rcl}
\nket{T(r,1)}^+&=&\Bigl(\prod_{i=-r}^{-1}\cG_i\Bigr)\,
\ket{\frac{1 - r}{t}}_{\rm top}\,,\\
\nket{T(r,1)}^-&=&\Bigl(\prod_{i=-r}^{-1}\cQ_i\Bigr)\,
\ket{-1 + \frac{1 + r}{t}}_{\rm top}\,,
\EA\quad r=1,2\ldots\EE
from which we construct $\nket{S(r,1)}^+$ as in \req{simple} and similarly
$\nket{S(r,1)}^-$.

In fact, the above recursive construction is {\it equivalent\/} to the
formula~\req{N2sing} as soon as one notices that~\req{N2sing} is arrived at
just by a successive application of the recursive prescriptions described in
this section, starting from the $\nket{T(r,1)}$ vectors, which would then
become precisely the product of modes of $\cG$ or $\cQ$ in the centre of the
formula~\req{N2sing} as we have noted.

\section{Concluding remarks\label{sec:Concl}}\lvm
The formula \req{N2sing} gives a general algebraic construction for the
topological singular vectors.   Comparison with the MFF construction shows
that the topological case involves more
complicated algebra. However, this is but a technical complication, since our
construction, just like the MFF one, is `direct' in that it does not require
solving equations and it allows one to derive singular vectors from a
generating expression by means of purely algebraic
manipulations.  The algebraic rules required in order to transform
the generating expression to the conventional form (a Verma module element)
consist of algebraic properties of the intertwining operators necessary for
manipulations in the generalized Verma modules and mappings between different
such modules.

The actual coincidence between the topological and affine $\SL2$ singular
vectors, observed in~\cite{[S-sing]}, should now be interpreted as the
statement that the isomorphism between the two classes of singular vectors is
supported by the two constructions -- MFF and ours -- {\it after\/} these
constructions are `continued' each in its own algebraic scheme.  When $r$,
$s$ and $t$ are chosen so that the formula for a topological singular vector
does not require a `continuation', the Kazama--Suzuki mapping can be applied
to it directly and takes it precisely into the corresponding MFF singular
vector, which in that case would be by itself (i.e.\ without any algebraic
manipulations either) a Verma module element.

A lesson that may be drawn from the existence of two generating
constructions is that general and at the same time
closed formulae for singular vectors should probably be expected to have an
`algebraically continued' form. This is almost certainly so for the $\N2$ $W_3$
singular vectors, but it is still not clear how general this feature may
actually be.  For the Virasoro algebra, the question is particularly acute,
since in the absence of any other operators except $L_n$, the algebra appears
to be too `poor' to allow `algebraically extended' constructions
that would allow for a monomial representation of singular vectors.  However,
an interesting fact is that the above derivation of topological singular
vectors involved a stage (Step~2 in the previous section) when what would
become a singular vector contained only $\cL$ and $\cH$ operators. That
`skeleton' form is therefore sufficient for the reconstruction of the entire
topological singular vector\footnote{Note that there is a certain `dualism' in
the roles played, on the one hand, by the fermionic generators $\cQ$ and $\cG$
and, on the other hand, by the bosonic ones $\cL$ and $\cH$. The generating
formula~\req{N2sing} is, heuristically, expressed in terms of $\cQ$ and $\cG$
alone, while a crucial step in the course of its rearrangement produces a
`skeleton' of the desired singular vector, involving only $\cL$ and $\cH$;
the `new' $\cG$ and $\cQ$ modes are then reconstructed from the
$\cL$-$\cH$-skeleton.}.  On the other hand, one can reduce the topological
singular vectors to those of the appropriate minimal matter
theory~\cite{[GS3],[Bea]}.  It would be interesting to see in detail how the
general construction for Virasoro singular vectors is recovered along these
lines, for instance by applying the reduction to the `skeleton' form.  More
generally, there exist a class of theories (minimal matter, $\SL2$, $\N2$,
parafermions, $\ldots$)  which are related by reductions (including the
hamiltonian reduction) or embeddings (such as embeddings of minimal matter
into $\N2$ \cite{[GS2],[BLNW]} or $\SL2$ \cite{[S-sing]} theories), and which
possess isomorphic (or `almost' isomorphic, say in relation 2:1) singular
vectors \cite{[ST3]}.  It would be interesting to see if different theories
from that class provide new reformulations of the generating construction for
essentially the same set of singular vectors.

It is also possible to extend the present construction to other $\N2$
singular vectors, which are built over non-chiral primary
states~\cite{[BFK]}. The construction for all singular vectors of the $\N2$
superconformal algebra is surprisingly similar to the one described in this
paper, and is given in~\cite{[ST4]}.

Let us note finally that, in view of the appearance of twisted $\N2$ algebra
in string theories~\cite{[GS2], [GS3],[BLNW]}, one may expect applications of
the present construction to physical states~\cite{[LZ]} with non-trivial
ghost numbers (the explicit form of MFF singular vectors has been used to
construct cohomology e.g.\ in~\cite{[HR]}).

\medskip

We would like to thank B.~Feigin  for useful discussions.  We are also
grateful to B.~Gato-Rivera, S.~Hwang, W.~Lerche, A.~Pogrebkov, M.A.~Soloviev,
K.~Stelle and Yu.~Zinoviev for useful comments.  This research was made
possible in part by Grant No.\ MQM300 from the International Science
Foundation and Government of Russian Federation, and by RFFI grant
94-02-06338-a.
\def\theequation{A.\arabic{equation}}

\subsection*{Appendix}\lvm Here we give two examples of the calculation
following the scheme described in section~\ref{sec:Rec}.
As a simple example, consider how the recursion procedure allows
one to derive the singular vector $\nket{T(1,3)}^-$ out of
\BE\ket{S(1,2)}^+=
\Bigl(-4t\cL_{-2} + 4t\cH_{-1}\cL_{-1}   +
4\cL_{-1}\cL_{-1} + 2t\cQ_{-1}\cG_{-1}
\Bigr)\ket{1}_{\rm top}
\label{step1}\EE
We now apply the steps described in Section~\ref{sec:Rec}. After the spectral
flow transform with parameter $1-2t$ we sandwich the expression in brackets
in \req{step1} between $q(-1, 2t - 2)\,\ldots\,q(2t-1,
-1)\nket{\frac{2}{t}-3}_{\rm top}$ and then drop the term
$2t\cQ_{2t-2}\cG_{-2t}$, in accordance with the `$q$'-version
of~\req{leftkill}. This leaves us with the `skeleton' inside the
intertwiners:
\BE\new\BA{rcl}
\ket{T(1,3)}^-&=&
q(-1, 2t - 2)\,\Bigl(
(-4 + 8t^2)\cH_{-2} -4t\cL_{-2} +
(4 - 12t + 8t^2)\cH_{-1}\cH_{-1}\\{}
{}&{}&{}+
(8 - 12t)\cL_{-1}\cH_{-1} +
4\cL_{-1}\cL_{-1}
\Bigr)\,q(2t-1,-1)\,\ket{\frac{2}{t}-3}_{\rm top}
\EA\EE
Here, we apply the `$q$'-analogue of the rules~\req{LHLeftg} (with
$g\leftrightarrow q$ and $\cG\leftrightarrow \cQ$) to move the modes of $\cL$
and $\cH$ through the left intertwiner $q(-1, 2t - 2)$.  Then the products of
the intertwiners rewrite as
$$\new\BA{c}q(-1, 2t - 2)\,q(2t-1,-1)= q(-1,-1)=\cQ_{-1}\,,\\
q(0, 2t - 2)\,q(2t-1,-1)= q(0,-1)=1\,.
\EA
$$
The result is
\BE\new\BA{rcl}
\ket{T(1,3)}^-&=&
\Bigl(
(-4 + 8t^2)\cH_{-2}\cQ_{-1} +
4t(-3 + 4t)\cH_{-1}\cQ_{-2} -4t\cL_{-2}\cQ_{-1} -12t\cL_{-1}\cQ_{-2}
\\{}&{}&{}+
(4 - 12t + 8t^2)\cH_{-1}\cH_{-1}\cQ_{-1} +
(8 - 12t)\cL_{-1}\cH_{-1}\cQ_{-1} +
4\cL_{-1}\cL_{-1}\cQ_{-1}\\{}
{}&{}&{}+ 4t(1 + 4t)\cQ_{-3}
\Bigr)\ket{\frac{2}{t}-3}_{\rm top}
\EA\EE
which is the sought singular vector.

\medskip

Consider now deriving the singular vector $\nket{T(1,5)}^+$ from
$\nket{S(1,4)}^-$.  The latter can be written as
\begin{eqnarray}
&&\ket{S(1,4)}^-=
\Bigl(
(-6 - 36 t - 66 t^2 - 36 t^3)\cH_{-4} +
(-6 t + 24 t^2 - 36 t^3)\cL_{-4} +
(t - 9 t^2 + 18 t^3)\cG_{-3}\cQ_{-1}\nonumber\\&&{} +
(-3 t^2 + 18 t^3)\cG_{-2}\cQ_{-2} +
(3 t^2 + 18 t^3)\cG_{-1}\cQ_{-3} +
(8 + 16 t - 32 t^2 - 48 t^3)\cH_{-3}\cH_{-1}\nonumber\\&&{} +
(3 + 6 t - 7 t^2 - 18 t^3)\cH_{-2}\cH_{-2} +
(-10 t + 39 t^2 - 36 t^3)\cL_{-3}\cH_{-1} +
(-10 t + 24 t^2)\cL_{-3}\cL_{-1}\nonumber\\&&{}+
(10 t + 12 t^2 - 18 t^3)\cL_{-2}\cH_{-2} + 9 t^2\cL_{-2}\cL_{-2} +
(8 + 28 t + 13 t^2 - 12 t^3)\cL_{-1}\cH_{-3}\nonumber\\&&{}+
(2 t - 14 t^2 + 18 t^3)\cG_{-2}\cH_{-1}\cQ_{-1} +
(2 t -\frac{21}{2} t^2)\cG_{-2}\cL_{-1}\cQ_{-1} +
(-3 t -\frac{3}{2} t^2 + 9 t^3)\cG_{-1}\cH_{-2}\cQ_{-1}\nonumber\\&&{}+
(-\frac{21}{2} t^2 + 18 t^3)\cG_{-1}\cH_{-1}\cQ_{-2} -
\frac{9}{2} t^2\cG_{-1}\cL_{-2}\cQ_{-1} -
\frac{21}{2} t^2\cG_{-1}\cL_{-1}\cQ_{-2}\label{work}\\&&{} -
(6 - 12 t - 22 t^2 + 36 t^3)\cH_{-2}\cH_{-1}\cH_{-1} -
(10 t - 30 t^2 + 18 t^3)\cL_{-2}\cH_{-1}\cH_{-1} +
3 t\cG_{-1}\cL_{-1}\cL_{-1}\cQ_{-1}\nonumber\\&&{}+
(-20 t + 30 t^2)\cL_{-2}\cL_{-1}\cH_{-1} -
10 t\cL_{-2}\cL_{-1}\cL_{-1} +
(-12 + 6 t + 43 t^2 - 18 t^3)\cL_{-1}\cH_{-2}\cH_{-1}\nonumber\\&&{}-
(6 + 6 t - 10 t^2)\cL_{-1}\cL_{-1}\cH_{-2} +
(3 t - 11 t^2 + 9 t^3)\cG_{-1}\cH_{-1}\cH_{-1}\cQ_{-1} +
(6 t - 11 t^2)\cG_{-1}\cL_{-1}\cH_{-1}\cQ_{-1}\nonumber\\&&{}+
(1 - 6 t + 11 t^2 - 6 t^3)\cH_{-1}\cH_{-1}\cH_{-1}\cH_{-1} +
(4 - 18 t + 22 t^2 - 6 t^3)\cL_{-1}\cH_{-1}\cH_{-1}\cH_{-1}\nonumber\\&&{}+
(6 - 18 t + 11 t^2)\cL_{-1}\cL_{-1}\cH_{-1}\cH_{-1} +
(4 - 6 t)\cL_{-1}\cL_{-1}\cL_{-1}\cH_{-1} +
\cL_{-1}\cL_{-1}\cL_{-1}\cL_{-1}
\Bigr)\ket{-4 +\frac{2}{t}}_{\rm top}\nonumber
\end{eqnarray} 
where we have chosen the ordering with $\cG$ modes on the left. Then, after
the spectral flow transform $\cG_n\mapsto\cG_{n+4t-1}$, the modes of $\cG$
will be immediately killed in accordance with \req{leftkill} when multiplied
from the left with the intertwiner \hbox{$g(-1,4t-2)$}. Performing in the
remaining ($\cQ$- and $\cG$- independent) terms the substitution
$\cL_n\mapsto\cL_n + (4t-1)\cH_n$, will then give the $\cL$-$\cH$-skeleton,
and the $\nket{T(1,5)}^+$ vector becomes $g(-1,4t-2)\cdot({\sl
skeleton\/})\cdot g(4t-1, -1)\nket{4}_{\rm top}$. Next, commuting the modes
of $\cL$ and $\cH$ from the skeleton through the intertwiner $g(-1,4t-2)$
according to the formulae~\req{LHLeftg}, and using \req{Un2g} and \req{Un1g},
we find the following terms which contain no $g(N, -1)$ factors of negative
length,
\begin{eqnarray}&&\Bigl(
(6 t + 83 t^2 + 380 t^3 + 576 t^4)\cG_{-5} -
(24 t + 144 t^2 + 264 t^3 + 144 t^4)\cH_{-4}\cG_{-1}\nonumber\\&&{}-
(72 t^2 + 256 t^3 + 192 t^4)\cH_{-3}\cG_{-2} -
(30 t^2 + 242 t^3 + 288 t^4)\cH_{-2}\cG_{-3} +
(25 t^2 - 60 t^3)\cL_{-3}\cG_{-2}\nonumber\\&&{}-
(22 t^2 + 228 t^3 + 576 t^4)\cH_{-1}\cG_{-4}-
(6 t - 24 t^2 + 36 t^3)\cL_{-4}\cG_{-1}  +
(32 t^2 + 96 t^3 + 72 t^4)\cH_{-2}\cH_{-2}\cG_{-1}\nonumber\\&&{}-
(10 t + 106 t^2 + 272 t^3)\cL_{-1}\cG_{-4} +
(72 t^2 + 256 t^3 + 192 t^4)\cH_{-3}\cH_{-1}\cG_{-1} -
(16 t^2 + 112 t^3)\cL_{-2}\cG_{-3}\nonumber\\&&{}+
(208 t^3 + 288 t^4)\cH_{-2}\cH_{-1}\cG_{-2} +
(46 t^3 + 288 t^4)\cH_{-1}\cH_{-1}\cG_{-3} +
(-25 t^2 + 60 t^3)\cL_{-3}\cH_{-1}\cG_{-1}\nonumber\\&&{}+
(-10 t + 24 t^2)\cL_{-3}\cL_{-1}\cG_{-1} +
(34 t^2 + 54 t^3)\cL_{-2}\cH_{-2}\cG_{-1} +
116 t^3\cL_{-2}\cH_{-1}\cG_{-2} + 9 t^2\cL_{-2}\cL_{-2}\cG_{-1}\nonumber\\&&{}+
50 t^2\cL_{-2}\cL_{-1}\cG_{-2} +
(30 t + 109 t^2 + 84 t^3)\cL_{-1}\cH_{-3}\cG_{-1} +
(95 t^2 + 138 t^3)\cL_{-1}\cH_{-2}\cG_{-2}\label{reg}\\&&{}+
(45 t^2 + 288 t^3)\cL_{-1}\cH_{-1}\cG_{-3} +
(10 t + 65 t^2)\cL_{-1}\cL_{-1}\cG_{-3} -
(104 t^3 + 144 t^4)\cH_{-2}\cH_{-1}\cH_{-1}\cG_{-1}\nonumber\\&&{} -
96 t^4\cH_{-1}\cH_{-1}\cH_{-1}\cG_{-2} -
58 t^3\cL_{-2}\cH_{-1}\cH_{-1}\cG_{-1} -
50 t^2\cL_{-2}\cL_{-1}\cH_{-1}\cG_{-1} -
10 t\cL_{-2}\cL_{-1}\cL_{-1}\cG_{-1}\nonumber\\&&{}-
(95 t^2 + 138 t^3)\cL_{-1}\cH_{-2}\cH_{-1}\cG_{-1} -
150 t^3\cL_{-1}\cH_{-1}\cH_{-1}\cG_{-2} -
(20 t + 30 t^2)\cL_{-1}\cL_{-1}\cH_{-2}\cG_{-1}\nonumber\\&&{}+
24 t^4\cH_{-1}\cH_{-1}\cH_{-1}\cH_{-1}\cG_{-1}+
50 t^3\cL_{-1}\cH_{-1}\cH_{-1}\cH_{-1}\cG_{-1}+
35 t^2\cL_{-1}\cL_{-1}\cH_{-1}\cH_{-1}\cG_{-1} \nonumber\\&&{}-
70 t^2\cL_{-1}\cL_{-1}\cH_{-1}\cG_{-2} -
10 t\cL_{-1}\cL_{-1}\cL_{-1}\cG_{-2}+
10 t\cL_{-1}\cL_{-1}\cL_{-1}\cH_{-1}\cG_{-1}  \nonumber\\&&\hfill{}+
\cL_{-1}\cL_{-1}\cL_{-1}\cL_{-1}\cG_{-1}\Bigr)\,\ket{4}_{\rm top}\nonumber
\end{eqnarray}
as well as 13 terms that do contain negative-length intertwiners:
\begin{eqnarray}
&&\Bigl(
(13 t^2 + 56 t^3)\cG_{-4}\cG_{-1}\,g(1, -1) +
(t^2 + 12 t^3)\cG_{-3}\cG_{-2}\,g(1, -1) -
(6 t + 26 t^2 - 52 t^3)\cG_{-3}\cG_{-1}\cG_0\,g(2, -1)\nonumber\\&&{}-
(4 t^2 + 20 t^3)\cH_{-2}\cG_{-2}\cG_{-1}\,g(1, -1) -
(3 t^2 + 40 t^3)\cH_{-1}\cG_{-3}\cG_{-1}\,g(1, -1)\nonumber\\&&{}-
(13 t^2 + 16 t^3)\cL_{-1}\cG_{-3}\cG_{-1}\,g(1, -1) +
(12 t + 8 t^2 - 32 t^3)\cG_{-2}\cG_{-1}\cG_0\cG_1\,g(3, -1)\nonumber\\&&{}+
(28 t^2 - 40 t^3)
\cH_{-1}\cG_{-2}\cG_{-1}\cG_0\,g(2, -1) +
12 t^3\cH_{-1}\cH_{-1}\cG_{-2}\cG_{-1}\,g(1, -1)\nonumber\\&&{}+
(10 t - 18 t^2 + 4 t^3)\cL_{-1}\cG_{-2}\cG_{-1}\cG_0\,g(2, -1) +
(5 t^2 + 12 t^3)
\cL_{-1}\cH_{-1}\cG_{-2}\cG_{-1}\,g(1, -1)\nonumber\\&&{}-
(2 t^2 + 4 t^3)\cL_{-2}\cG_{-2}\cG_{-1}\,g(1, -1) +
5 t^2\cL_{-1}\cL_{-1}\cG_{-2}\cG_{-1}\,g(1, -1)\Bigr)\,\ket{4}_{\rm top}
\label{weird15}\end{eqnarray}
To `resolve' the negative-length intertwiners, we proceed as in Sect.~7.
First, $\cG_{-2}\cG_{-1}\cG_0\cG_1\,g(3, -1)\,\nket{4}_{\rm top}$ rewrites as
${t\over4(4 t - 3)}\,\cG_{-2}\cG_{-1}\cG_0\cG_1\cQ_{-2}\,g(2,
-1)\,\nket{4}_{\rm top}$. Then,
commuting $\cG_1$ through $\cQ_{-2}$, we would get
\ $\cG_1\,g(2, -1)\,\nket{4}_{\rm top}=g(1, -1)\,\nket{4}_{\rm top}$. To all
the remaining terms that contain $g(2,-1)\,\nket4_{\rm top}$ we apply the
formula \req{Fill1} again, and thus \req{weird15} becomes
\begin{eqnarray}
&&\Bigl(
(8 t^2 + 36 t^3)\cG_{-5} + (7 t^2 + 36 t^3)\cH_{-4}\cG_{-1} -
(4 t^2 + 20 t^3)\cH_{-3}\cG_{-2} -
(3 t^2 + 16 t^3)\cH_{-2}\cG_{-3}\nonumber\\&&{} -
12 t^3\cH_{-1}\cG_{-4} +
(-5 t^2 + 20 t^3)\cL_{-4}\cG_{-1} + (3 t^2 - 4 t^3)\cL_{-3}\cG_{-2} -
(3 t^2 + 16 t^3)\cL_{-2}\cG_{-3}\nonumber\\&&{}-
5 t^2\cL_{-1}\cG_{-4} -
12 t^3\cH_{-3}\cH_{-1}\cG_{-1} + 12 t^3\cH_{-2}\cH_{-1}\cG_{-2} -
12 t^3\cL_{-3}\cH_{-1}\cG_{-1} - 5 t^2\cL_{-3}\cL_{-1}\cG_{-1}\nonumber\\&&{}+
12 t^3\cL_{-2}\cH_{-1}\cG_{-2} + 5 t^2\cL_{-2}\cL_{-1}\cG_{-2} -
5 t^2\cL_{-1}\cH_{-3}\cG_{-1} + 5 t^2\cL_{-1}\cH_{-2}\cG_{-2} -
(t^2 + 2 t^3)\cQ_{-2}\cG_{-2}\cG_{-1}\nonumber\\&&{}-
(\frac{3}{2} t^2 + 8 t^3)\cQ_{-1}\cG_{-3}\cG_{-1} +
6 t^3\cH_{-1}\cQ_{-1}\cG_{-2}\cG_{-1} +
\frac{5}{2} t^2\cL_{-1}\cQ_{-1}\cG_{-2}\cG_{-1}\Bigr)\,\ket{4}_{\rm top}
\end{eqnarray}
Adding these terms with \req{reg} does give the singular vector
$\nket{T(1,5)}^+$.


\newpage
\small
\begin{thebibliography}{99}
\parindent=0pt
\parskip=-2pt
\bibitem{[BPZ]} A.A.~Belavin, A.M.~Polyakov, and A.B.~Zamolodchikov,
\NPB241 (1984) 333.

\bibitem{[FF]}B.L.~Feigin and D.B.~Fuchs, {\sl Representations of the
Virasoro algebra\/}, in: {\it Representations of infinite-dimension\-al Lie
groups and algebras\/}, N.-Y., Gordon and Breach, 1986.

\bibitem{[KK]}V.G.~Ka\v c and D.A.~Kazhdan, Adv.\ Math.\ 34 (1979) 97.

\bibitem{[FQS]} D.~Friedan, Z.~Qiu, and S.~Shenker, Phys.\ Rev.\ Lett. 52
(1984) 1575.

\bibitem{[BFK]} W.~Boucher, D.~Friedan, and A.~Kent, \PLB172 (1986) 316.

\bibitem{[K]} E.B.~Kiritsis, Phys.\ Rev.\ D36 (1987) 3048.

\bibitem{[MFF]} F.G.~Malikov, B.L.~Feigin, and D.B.~Fuchs, Funk.\ An.\
Prilozh.\ 20 N2 (1986) 25.

\bibitem{[BSa]} L.~Benoit and Y.~Saint-Aubin, \PLB215 (1988) 517.

\bibitem{[K1]} A.~Kent, \PLB273 (1991) 56.

\bibitem{[BdFIZ]} M.~Bauer, P.~di~Francesco, C.~Itzykson, and J.-B.~Zuber,
\NPB362 (1991) 515.

\bibitem{[GP]} A.Ch.~Ganchev and V.B.~Petkova, \PLB293 (1992) 56; \PLB318
(1993) 77;\\
P.~Furlan, A.Ch.~Ganchev, and V.B.~Petkova, \NPB431 (1994) 622-666

\bibitem{[BSa1]}L.~Benoit and Y.~Saint-Aubin, \IJMPA7 (1992) 3032.

\bibitem{[W1]}G.M.T.~Watts, \NPB407 (1993) 213.

\bibitem{[HZZ]}C.-S.~Huang, D.-H.~Zhang and Q.-R.~Zheng, \NPB389 (1993) 81.

\bibitem{[BW]} P.~Bowcock and G.M.T.~Watts, \PLB297 (1992) 282.

\bibitem{[W]} G.M.T.~Watts, \NPB407 (1993) 213.

\bibitem{[BS]} M.~Bauer and N.~Sochen, Comm.\ Math.\ Phys.\ 152 (1993) 127.

\bibitem{[S-sing]} A.M.~Semikhatov, \MPLA9 (1994) 1867.

\bibitem{[B]} Z.~Bajnok, \PLB320 (1994) 36; \PLB329 (1994) 225.

\bibitem{[Doerr]}M.~D\"orrzapf, \IJMPA10 (1995) 2143.

\bibitem{[ST3]} A.M.~Semikhatov and I.Yu.~Tipunin, {\it $N=2$, $s\ell(2)$,
and related conformal models: a Universal Realization and singular
vectors\/}, in preparation.

\bibitem{[Bea]}B.~Gato-Rivera and J.-I.~Rosado, \PLB346 (1995) 63; {\sl
Spectral flows and twisted topological theories\/}, {\tt hep-th/9504056}.

\bibitem{[FM]}B.~Feigin and F.~Malikov, {\sl Integral intertwining operators
and complex powers of differential ($q$-difference) operators\/}.

\bibitem{[SS]}A.~Schwimmer and N.~Seiberg, \PLB184 (1987) 191.

\bibitem{[LVW]}W.~Lerche, C.~Vafa, and N.P.~Warner, \NPB324 (1989) 427.

\bibitem{[EY]}T.~Eguchi and S.-K.~Yang, \MPLA4 (1990) 1653.

\bibitem{[W-top]}E.~Witten, Commun.\ Math.\ Phys.\ 118 (1988) 411; \NPB 340
(1990) 281.

\bibitem{[GS3]}B.~Gato-Rivera and A.M.~Semikhatov, \NPB408 (1993) 133.

\bibitem{[KS]}Y.~Kazama and H.~Suzuki, \NPB321 (1989) 232.

\bibitem{[Lerche-pr]}W.~Lerche, private communication.

\bibitem{[ST2r]} A.M.~Semikhatov and I.Yu.~Tipunin, ZhETF Lett.\ 63
(1996) 129.

\bibitem{[GS2]} B.~Gato-Rivera and A.M.~Semikhatov, \PLB288 (1992) 38.

\bibitem{[BLNW]}M.~Bershadsky, W.~Lerche, D.~Nemeschansky, and N.~P.~Warner,
\NPB401 (1993) 304.

\bibitem{[LZ]}B.H.~Lian and G.J.~Zuckerman, \PLB254 (1991) 417.

\bibitem{[HR]}S.~Hwang and H.~Rhedin, \PLB350 (1995) 38.

\bibitem{[ST4]} A.M.~Semikhatov and I.Yu.~Tipunin, {\it All Singular Vectors
of the $N=2$ Superconformal Algebra via the Algebraic Continuation
Approach\/}, {\tt hep-th/9604176}.

\end{thebibliography}
\end{document}